\documentclass[apj]{emulateapj_rtx4}
\usepackage{mathrsfs}

\usepackage{mathptmx}
\usepackage{epsfig,natbib}

\slugcomment{Accepted by Astrophysical Journal Supplement Series}

\shortauthors{Wu et al.} \shorttitle{Evolution in starburst galaxies
and AGNs}

\def\aj{\rm{AJ}}                   
\def\araa{\rm{ARA\&A}}             
\def\apj{\rm {ApJ}}                
\def\apjl{\rm{ApJ}}                
\def\apjs{\rm{ApJS}}               
\def\aap{\rm{A\&A}}                
\def\aaps{\rm{A\&AS}}              
\def\mnras{\rm{MNRAS}}             
\def\pasp{\rm{PASP}}           
\def\starlight{\textsc{starlight}}

\def\cd{c$\!\!\!\hskip 0.75pt$\raise 0.2pt \hbox{\symbol{24}}}

\def\Msol{\ifmmode{\rm M}_{\mathord\odot}\else M$_{\mathord\odot}$\fi}
\def\Mbh{\ifmmode{M_{\rm BH}}\else{$M_{\rm BH}$}\fi}
\def\REdd{\ifmmode{{\cal R}_{\rm Edd}}\else{${\cal R}_{\rm Edd}$}\fi}
\def\nLn{\ifmmode{\nu L_{\nu}}\else{$\nu L_{\nu}$}\fi}

\def\ls{\lower 2pt \hbox{$\;\scriptscriptstyle \buildrel<\over\sim\;$}}
\def\gs{\lower 2pt \hbox{$\;\scriptscriptstyle \buildrel>\over\sim\;$}}

\def\kms{\ifmmode{{\rm km~s^{-1}}}\else{km~s$^{-1}$}\fi}
\def\ergs{\ifmmode{{\rm erg~s^{-1}}}\else{erg~s$^{-1}$}\fi}
\def\m#1{\ifmmode{^{-#1}}\else{$^{-#1}$}\fi}
\def\asec{\ifmmode{^{\prime\prime}}\else{$^{\prime\prime}$}\fi}
\def\asecb{\ifmmode{^{\prime\prime\!\!\!}}\else{$^{\prime\prime\!\!\!}$}\fi}
\def\asecp{\ifmmode{^{\prime\prime\!\!\!}.}\else{$^{\prime\prime\!\!\!}$.}\fi}
\def\deg{\ifmmode{^{\circ}}\else{$^{\circ}$}\fi}
\def\degp{\ifmmode{^{\circ\!\!\!}.}\else{$^{\circ\!\!\!}$.}\fi}
\def\aox{\ifmmode{\alpha_{\rm ox}}\else{$\alpha_{\rm ox}$}\fi}
\def\zbol{\ifmmode{\kappa_{\rm 2-10\; keV}}\else{$\kappa_{\rm 2-10\; keV}$}\fi}

\def\ten#1{\ifmmode{\times 10^{#1}}\else{$\times 10^{#1}$}\fi}
\def\tten#1{\ifmmode{\times 10^{#1}}\else{$\times 10^{#1}$}\fi}
\def\nten#1#2{\ifmmode{#1\times 10^{#2}}\else{$#1\times 10^{#2}$}\fi}

\newcounter{species}
\def\ion#1#2{\setcounter{species}{#2}#1$\;${\sc\roman{species}}\relax}

\def\l{$\lambda$}

\def\a{$\alpha$}
\def\b{$\beta$}

\def\spitzer{{\it Spitzer}}
\def\MPA-JHU{{\it MPA-JHU DR7}}

\def\spi¡¯s{{\it Spitzer's}}

\def\HII  {\ion{H}{2}}
\def\NaI  {\ion{Na}{1}}
\def\NeI {[\ion{Ne}{3}]}
\def\HeI  {\ion{He}{1}}

\def\NII  {[\ion{N}{2}]}
\def\NeII {[\ion{Ne}{2}]}

\def\NeV  {[\ion{Ne}{5}]}
\def\OI   {[\ion{O}{1}]}
\def\OII  {[\ion{O}{2}]}
\def\oiii {[\ion{O}{3}]}
\def\oiv  {[\ion{O}{4}]}
\def\SII  {[\ion{S}{2}]}
\def\Siii {[\ion{S}{3}]}
\def\Si ii {[\ion{Si}{2}]}

\begin{document}

\title{The Diagnostics and Possible Evolution in Active Galactic Nuclei
  Associated With Starburst Galaxies}

\author{Yu-Zhong Wu\altaffilmark{1,2,3}
, Yong-Heng Zhao\altaffilmark{1,3}, and Xian-Min
Meng\altaffilmark{1}}

\altaffiltext{1}{National Astronomical Observatories, Chinese
Academy of Sciences, 20A Datun Road, Beijing 100012, China;
yzwu@nao.cas.cn}

\altaffiltext{2}{Graduate University of Chinese Academy of Sciences,
19A Yuanquan Road, Beijing 100049, China.}

\altaffiltext{3}{Key Laboratory of Optical Astronomy, National
Astronomical Observatories, Chinese Academy of Sciences, Beijing
100012, China.}

\begin{abstract}
We present a large sample which contains 45 Seyfert 1 galaxies
(Sy1s), 46 hidden broad-line region (HBLR) Seyfert 2 galaxies
(Sy2s), 57 non-HBLR Sy2s, and 22 starburst galaxies to distinguish
their properties and to seek the possible evolution of active
galactic nuclei (AGNs) and starburst galaxies. We show that (1)
using a plot of \oiii~\l 5007/H\a~versus \NII~\l 6584/H\a~of
standard optical spectral diagnostic diagrams, we find that a
equation can separate well non-HBLR Sy2s and HBLR Sy2s; (2) the
emission-line ratios and both the combination of the ratios and
polycyclic aromatic hydrocarbon (PAH) strength are utilized
effectively to separate starburst galaxies and Seyfert galaxies; (3)
we compare a number of quantities from the data and confirm well the
separations with statistics; (4) on the basis of statistics, we
suggest that HBLR Sy2s may be the counterparts of Sy1s at edge-on
orientation. In addition, we discuss the possibility of starburst
galaxies evolving to non-HBLR Sy2s and HBLR Sy2s and then evolving
to Sy1s based on the statistical analysis.

\end{abstract}


\section{Introduction}\label{S:intro}

To constrain the galaxy formation and evolution process, it is
essential to understand the nuclear activity in nearby galaxies
(Constantin et al. 2009). As galaxies evolve, they usually
experience the phases of their star formation rates increasing
substantially, which is often ascribed to large-scale perturbations
in their structure, such as, interactions triggering the
formation of bars, resulting in the accumulation of large numbers of
gas in their nuclei (Bernard-Salas et al. 2009). For the complexity
and importance of the starburst and nuclear activities in the
nuclear region, the study of starburst-active galactic nucleus (AGN)
connection is a crucial component of modern astrophysics. Based on
the results of X-ray imaging and spectroscopic analysis of a Seyfert 2
galaxy (Sy2) sample, starbursts and AGNs are regarded as part
of a general evolutionary sequence (Levenson et al. 2001).

Some of the fundamental study for the evolution can date back to the
1980s. Close companions or part of interacting systems were found in
the Seyfert galaxies and luminous $Infrared~Astronomical~Satellite
~(IRAS)$ galaxies ($L_{\rm FIR}>10^{11}L_{\sun}$; Mirabel $\&$
Wilson 1984; Sanders et al. 1986). In addition, many results
indicated that the galaxy interactions play a critical role in
boosting not only the far-infrared luminosity but also the $L_{\rm
FIR}/M(\rm H_{2})$ ratio (Sanders $\&$ Mirabel 1985; Young et al.
1986), and in the case of strong interactions or mergers, they
significantly increase the total supply of molecular gas fueling the
starburst. The merger-driven evolutionary sequence initially
proposed by Sanders et al. (1988) suggests that the merger of two
gas-rich disk galaxies can eventually form an elliptical galaxy.

Based on different methods, some studies seem to conclude that there
are indeed some evolutionary processes in some galaxies. Investigating some relations
for a sample of 35 Seyfert 2 nuclei, Storchi-Bergmann et al. (2001)
found that the star formation activity nearby Seyfert 2 nuclei is
linked to interactions and suggested an evolutionary scenario which
the starburst dies down with time while the composite Seyfert
2$~+~$starburst nucleus evolves to a ``pure" Seyfert 2 nucleus with
an old stellar population. Using the bright infrared galaxies
(BIRGs) sample and related studies, Koulouridis et al. (2006)
suggested that molecular clouds moving toward the galactic center
are drived by close interactions, which can both trigger starburst
activity and obscure the nucleus. Under the similar environment
found for the BIRGs and Sy2s, Krongold et al. (2002) suggested an
evolutionary link between starbursts, Sy2s, and Sy1s and proposed the
evolutionary scenrio for AGNs: interaction
$\rightarrow$ starburst$\rightarrow$ Seyfert 2 $\leftrightarrow$
Seyfert 1.

 Oliva et al. (1999c) found that old and powerful
 starbursts are absent in Seyfert 1 galaxies while relatively
  common in obscured (type 2) AGNs. Some studies also show that the majority of type 2 AGNs appear to
have experienced significant starbursts (Kauffmann et al. 2003a, b).
As suggested in Buchanan et al. (2006) that if star formation is
strong in Sy2s, then Seyfert 2 galaxies may evolve to Seyfert 1
galaxies (Tommasin et al. 2010). A similar scenario that Seyfert 2
galaxies are the transitional stages of \HII/starburst galaxies
evolving into Seyfert 1 galaxies was proposed: \HII $\rightarrow $
Seyfert 2 $\rightarrow $ Seyfert 1 (Hunt \& Malkan 1999; Levenson et
al. 2001; Tommasin et al. 2010).

The goal of this study is to explore whether both Seyfert and
starburst galaxies can be distinguished by optical and infrared
spectral diagnostics; then we investigate and discuss a possibility
that whether a evolutionary sequence of starburst
galaxies$\rightarrow$ non-hidden broad line region (HBLR) Sy2s
$\rightarrow$ HBLR Sy2s $\rightarrow$ Sy1s could be invoked.

This paper is organized as follows. In Section 2, we show the sample
of galaxies and their basic properties in detail and provide details
of how we selected the data. Section 3 displays the different
diagnostics through various quantities. In Section 4, we distinguish
between starburst galaxies, non-HBLR Sy2s, HBLR Sy2s, and
 Sy1s with a number of quantities, namely the optical emission-line ratio,
$f_{60}/f_{25}$ ratio, several mid-infrared line ratio diagnostics,
and $6.2~\mu m~$polycyclic aromatic hydrocarbon (PAH) equivalent
width (EW) and suggest the possible evolutionary scenario. In
Section 5, we discuss the separation among them with the optical
 and mid-infrared spectral diagnostics, respectively and the
 possibility of the evolution sequence from starburst
galaxies, non-HBLR Sy2s, and HBLR Sy2s to Sy1s. In Section 6, we
summarize the results and the conclusions.

\def\oiii{\ifmmode [O {\sc iii}] \else [O {\sc iii}]\ \fi}
\def\Neii{\ifmmode [O {\sc ii}] \else [Ne {\sc ii}]\ \fi}
\def\Neiii{\ifmmode [O {\sc iii}] \else [Ne {\sc iii}]\ \fi}
\def\Nev{\ifmmode [O {\sc v}] \else [Ne {\sc v}]\ \fi}
\def\oiv{\ifmmode [O {\sc iv}] \else [O {\sc iv}]\ \fi}
\def\Feii{\ifmmode [Fe {\sc ii}] \else [Fe {\sc ii}]\ \fi}
\def\Siii{\ifmmode [S {\sc iii}] \else [S {\sc iii}]\ \fi}
\def\Siiv{\ifmmode [O {\sc ii}] \else [Si {\sc ii}]\ \fi}

\newcommand{\OIII}{[\rm O~ \sc{III}]}
\newcommand{\LOIII}{L_{\rm [O~ \sc{III}]}}



\begin{table*}
\caption{\small The Sy1 sample}
\setlength{\tabcolsep}{6.0pt}
\renewcommand{\arraystretch}{1.0}

\begin{tabular}{lccccccccccl} \hline \hline

name  & $f_{25}$ & $f_{60}$ & \Neii & \Nev &\Neiii& \oiv & \Feii
& \Siii & \Si ii &$ \rm EW$ &Reference \\
&  & &$12.81~\mu m $&$14.32~\mu m$&$15.56~\mu m$&$25.89~\mu
m$&$25.99~\mu m$&$33.48~\mu m$&$34.28~\mu m$&$6.2~\mu m$ \\
      (1) & (2) &(3) & (4) & (5) & (6)&(7)& (8) & (9)& (10)& (11)&(12) \\
\hline

ESO 012-G21 & 0.25 & 1.45 & 11.95 &3.19     & 6.42  & 15.98  &...    & 11.17&26.8   &0.278    &27,13,13,19\\
ESO 0141-G55& 0.46 & 0.47 &  2.24 &  2.25   & 5.62  &  7.26  & ...   &5.45  & 8.85  &...      &27,13,13\\
F03450+0055 & 0.39 & 0.87 & 1.09  &$<$1.48  & 1.82  & 2.52   & ...   & 1.54 &$<$4.5 &$<$0.103 &27,14,14,19\\
F05563-3820 & 0.77 & 0.38 & 3.89  &  2.54   & 4.82  & 5.31   & ...  &$<$1.84&2.89   &....     &27,14,14\\
F13349+2438 & 0.72 & 0.85 & ...   &  ...    & ...   & ...    & ...   &...   & ...   &....     &27\\
F15091-2107 & 0.97 & 1.60 & 11.52 &  8.48   & 16.29 &  31.03 & ...   &12.26 & 12.27 &....     &27,13,13\\
IC 4329A    & 2.26 & 2.15 & 27.6  &  29.3   & 57.0  &  117.0 & ...   &16.0  &32.5   &$<$0.02  &27,14,14,19\\
I ZW 1      & 1.17 & 2.24 & 1.90  & 5.50    & 4.50  &  2.70 &$<$1.60&$<$2.00&  4.40 &0.01     &27,15,15,15,18\\
MCG-2-33-34 & 0.65 & 1.23 &7.37   &6.75     & 15.93 & 81.93  & ...   &20.42 & 29.46 &$<$0.304 &27,13,13,19\\
MCG-3-7-11  & 0.35 & 1.45 &  ...  & ...     & ...   & ...    & ...   & ...  &  ...  & ...     &27\\
MCG-5-13-17 & 0.57 & 1.28 & 11.0  &  ...    & 13.0  & 12.0   & ...   &15.0  & ...   &0.20     &27,16,16,19\\
MCG-6-30-15 & 0.97 & 1.39 & 4.98  & 5.01    & 5.88  & 26.0   & ...   &6.51  & 9.26  &$<$0.063 &27,14,14,19\\
Mrk 6       & 0.73 & 1.25 & 28.0  & 9.39    & 49.34 & 48.24  & ...   &14.09 &36.40  &$<$0.097 &27,13,13,19\\
Mrk 9       & 0.44 & 0.77 & 3.23  & 2.21    & 1.90  &  5.55  & ...   &3.94  &7.32   &$<$0.142 &1,13,13,19\\
Mrk 79      & 0.73 & 1.55 &   10.2& 6.55    & 19.6  &  42.0  & ...   & 14.0 & 30.5  &$<$0.039 &27,14,14,19\\
Mrk 205     & 0.08 & 0.29 &  ...  &  ...    & ...   &   ...  &  ...  & ...  & ...   &...      &27\\
Mrk 231     & 8.80 & 35.4 &  18.50&$<$6.20  &$<$5.60&$<$10.00& ... &$<$32.50&  ...  &0.011    &27,18,18,19\\
Mrk 279     & 0.50 & 1.58 &  ...  &3.28     & ...   & 10.82  &....   & ...  &  .... &...      &27,21\\
Mrk 335     & 0.45 & 0.35 &$<$7.0 &$<$5.0   &$<$8.0 &   13.0 &$<$3.0 &$<$1.21&$<$1.45&$<$0.074&27,12,12,14,19\\
Mrk 509     & 0.73 & 1.39 &  14.0 & 4.74    & 14.5  &   27.5 & ...   &7.41   & 14.5 &0.042    &27,14,14,19\\
Mrk 618     & 0.85 & 2.70 & 16.4  &  3.89   & 5.38  &  10.2  & ...   &6.35   &78.0  &....     &27,14,14\\
Mrk 704     & 0.60 & 0.36 &$<$3.30&  3.93   & 5.63  & 11.8   & ...   &$<$4.30&...   &$<$0.071 &27,14,14,19\\
Mrk 817     & 1.42 & 2.33 & 3.83  &   1.86  & 4.58  &  6.53  & ...   &$<$3.21&$<$0.45&$<$0.109&27,13,13,19\\
Mrk 841     & 0.47 & 0.46 & 3.60  &  8.0    & 11.0  &  24.0  &$<$1.0 &4.50   &  5.50&...      &1,15,15,15\\
Mrk 993     &$<$0.13& 0.30& ...   & ...     & ...   &  ...   & ...   &...    &...   & ...     &1\\
Mrk 1239    & 1.21 & 1.41 &9.4    & 3.4     & 9.38  & 15.6   & ...   & 9.09  & 10.5 &$<$0.029 &27,14,14,19\\
NGC863      & 0.22 & 0.49 &  ...  &1.01     & ...   &  3.39  & ...   &...    &...   &....     &27,21\\
NGC931      & 1.42 & 2.80 & 5.47  &   14.30 & 15.41 &  42.60 & ...   &11.97  &13.72 &$<$0.06  &27,13,13,19\\
NGC1566$^a$ & 3.07 &23.12 &21.09  &   1.19  &11.53  &  7.50  & 1.56  & 9.03  &18.37 &  0.223  &27,23,23,23,19\\
NGC3080     &$<$0.15& 0.35&  ...  &    ...  & ...   &  ...   & ...   &...    &...   &...      &27\\
NGC3227     & 2.04 &9.01  &  ...  & 23.13   &  ...  & 65.37  & ...  &  19.00 & ...  & 0.138   &30,21,16,19\\
NGC3516     & 0.96 & 2.09 & 8.07  &  7.88   & 17.72 & 46.92  & ...   & 9.52  &22.14 &$<$0.061 &27,13,13,19\\
NGC4051     & 2.28 & 10.62& 21.2  &  10.7   & 17.1  & 94.6   & ...   &38.8   &39.6  &0.079    &27,14,14,19\\
NGC4151     & 5.04 & 5.64 & 118.0 &  55.0   & 207.0 & 203.0  & 4.0   &81.0   &156.0 &$<$0.011 &27,12,12,12,19\\
NGC4235     & 0.28 & 0.65 &  ...  &...      & ...   &   ...  & ...   &...    &...   &...      &27\\
NGC4253     & 1.47 & 3.89 &  23.3 &  21.0   & 24.1  & 46.1   & ...   &21.4   &15.5  &0.036    &27,14,14,19\\
NGC4593     & 0.96 & 3.43 & 7.34  &  3.09   & 8.13  &  33.3  & ...   &19.1   &32.2  &0.044    &27,14,14,19\\
NGC5548     & 0.81 & 1.07 & 8.47  &   5.4   & 7.27  &  17.5  & ...   &$<$4.14&12.46 &0.018    &27,14,14,19\\
NGC5940     & 0.11 & 0.74 &  ...  &  ...    & ...   &   ...  & ...   & ...   & ...  &...      &27\\
NGC6104     & 0.16 & 0.76 &  ...  &   ...   & ...   &   ...  & ...   &...    &...   &...      &27\\
NGC6860     & 0.31 & 0.96 &  5.60 &   2.85  & 6.65  &   12.1 & ...   &7.93   & 10.4 &0.084    &27,14,14,19\\
NGC7213     & 0.81 & 2.70 &  25.7 &$<$1.85  & 12.0  &$<$13.5 & ...   &6.97   & 15.7 &0.022    &27,14,14,19\\
NGC7469     & 6.04 & 28.57& 226.0 &$<$15.0  & 22.0  &34.0    & 6.0   &104.0  &196.0 &0.208    &27,12,12,12,18\\
NGC7603     &$<$0.24& 0.85& 11.0  &   ...   & 7.0   &   7.0  & ...   &10.0   &...   &0.056    &1,16,16,19\\
UGC524      & 0.17 & 0.94 &  ...  &  ...    & ...   &  ...   & ...   &...    &...   &....     &27\\
\hline

\end{tabular}

{ \noindent \vglue 0.5cm {\sc Notes}: Column 1: Source name. Columns
2 and 3: Infrared flux (in janskys) for 25 and 60 $\mu m$. Columns
$4-10$: Flux ($10^{-21}$W $\rm cm^{-2}$) for \Neii~$12.81~\mu m$,
\Nev~$14.32~\mu m$, \oiv~$25.89~\mu m$, \Feii~$25.99~\mu m$,
\Siii~$33.48~\mu m$, and \Si ii ~$34.28~\mu m$, respectively. Column
11: Equivalent width in $\mu m$. Column 12: References (for columns.
2-3, 4-7, 8, 9-10 and 11, respectively). The fluxes listed of
reference 25 between the two high-resolution modules, short-high and
long-high, are not to be directly compared because of the different
aperture used (Bernard-Salas et al. 2009).

$^a$ Fluxes (except 25 and 60$\mu$m) are listed in units of $\rm
10^{-9} W~m^{-2}~sr^{-1}$.

This table has the same References as the table 2.}



\end{table*}






\def\oiii{\ifmmode [O {\sc iii}] \else [O {\sc iii}]\ \fi}
\def\Neii{\ifmmode [Ne {\sc ii}] \else [Ne {\sc ii}]\ \fi}
\def\Neiii{\ifmmode [O {\sc iii}] \else [Ne {\sc iii}]\ \fi}
\def\Nev{\ifmmode [O {\sc v}] \else [Ne {\sc v}]\ \fi}
\def\oiv{\ifmmode [O {\sc iv}] \else [O {\sc iv}]\ \fi}
\def\Feii{\ifmmode [Fe {\sc ii}] \else [Fe {\sc ii}]\ \fi}
\def\Siii{\ifmmode [S {\sc iii}] \else [S {\sc iii}]\ \fi}
\def\Si ii{\ifmmode [Si {\sc ii}] \else [Si {\sc ii}]\ \fi}



\begin{table*}
\caption{\small The HBLR Sy2 sample}
\setlength{\tabcolsep}{2.5pt}
\renewcommand{\arraystretch}{1.2}
\begin{tabular}{lccccccccccccl} \hline \hline
name  & $ \rm \frac{[O~III ] \lambda5007}{H\beta}$ & $\frac{[\rm
N~II]\lambda6584}{H\alpha}$ & $f_{25}$ & $f_{60}$ & \Neii & \Nev
&\Neiii& \oiv &   \Feii$^{a}$
&  \Siii & \Si ii &$ \rm EW$ &Reference \\
&  &  &  & &$12.81~\mu m $&$14.32~\mu m$&$15.56~\mu m$&$25.89~\mu
m$&$25.99~\mu m$&$33.48~\mu m$&$34.28~\mu m$&$6.2~\mu m$ \\
      (1) & (2) &(3) & (4) & (5) & (6)&(7)& (8) & (9)& (10)& (11)&(12) &(13)&(14) \\
\hline

Circinus      &   10.25& 1.15  & 68.44 &248.7 &  900.0 & 317.0  & 335.0 & 679.0& 59.0  &932.0 &1510.0& ...    &20,1,12,12,12\\
ESO273-IG04   &   ...  & ...   & 1.72  & 4.76 &   ...  & ...    & ...   &  ... & ...   & ...  & ...  & ...    &1\\
F00521-7054   &   ...  & ...   & 0.80  &1.02  & 5.80   & 5.78   & 8.13  &  8.63& ...   & 3.75 &...   & ...    &1,13,13\\
F01475-0740   &   5.21 & 0.49  & 0.84  & 1.10 &  13.7  &  6.38  & 9.95  &  6.49& ...   &3.12 &$<$6.13&$<$0.177&4,27,14,14,19\\
F04385-0828   &   ...  & ...   & 1.70  & 2.91 & 13.9   & 2.28   & 7.06  & 8.56 & ... &$<$5.23&$<$5.36&$<$0.058&27,14,14,19\\
F05189-2524   &   ...  & ...   & 3.41  & 13.27&  16.0  & 19.0   & 18.0  & 28.0 & 3.50 &$<$11.0&$<$16.0& 0.037 &27,15,15,15,19\\
F11057-1131   &   ...  & ...   & 0.32  & 0.77 &  ...   &  ...   & ...   & ...  & ...  &...    & ...  & ...    &1\\
F15480-0344   &    ... & ...   & 0.72  & 1.09 &  5.57  & 6.08   & 9.35  & 35.0 & ...  &5.20   & 5.13 &$<$0.19 &27,14,14,19\\
F17345+1124   &   ...  & ...   & 0.20  & 0.48 &  ...   &   ...  & ...   &  ... & ...  &...    & ...  & ...    &1\\
F18325-5926   &  ...   & ...   & 1.39  & 3.23 & ...    &  ...   & ...   &  ... & ...  & ...   & ...  & ...    &1\\
F20460+1925   &  ...   & ...   & 0.53  & 0.88 &  ...   &  ...   & ...   &  ... & ...  &...    & ...  & ...    &1\\
F22017+0319   &  ...   & ...   & 0.59  & 1.31 &   5.95 &   8.33 & 14.07 & 29.04& ...  &9.33   & ...  & ...    &27,13,13\\
F23060+0505   &  ...   & ...   & 0.43  & 1.15 &  ...   &  ...   & ...   & ...  & ...  &...    & ...  & ...    &1\\
IC 3639       &  9.55  & 0.78  & 2.54  & 8.90 &  45.15 &  11.15 & 27.00 & 21.21& ...  &32.80  &44.99 & 0.185  &10,28,13,13,19\\
IC 5063       &  8.03  & 0.60  & 3.95  & 5.79 &  26.7  &  30.3  &66.3   &114.0 & 5.0  & 31.0  &52.7  & 0.011  &6,27,14,12,14,19\\
MCG-2-8-39    &  ...   & ...   & 0.48  & 0.51 &  3.86  &   6.59 & 9.79  & 14.4 &  ... &$<$4.31& ...  &$<$0.128&1,14,14,19\\
MCG-3-34-64   &  ...   & ...   & 2.88  & 6.22 &  ...   &  ...   & ...   &  ... & ...  &...    & ...  & ...    &27\\
MCG-3-58-7    &  ...   & ...   & 0.98  & 2.60 &    8.52& 6.63   & 9.29  & 8.80 & ...  &$<$2.84& 11.7 & 0.074  &27,14,14,19\\
MCG-5-23-16   &  0.40  & 0.78  & ...   & ...  &   ...  &   ...  & ...   &  ... & ...   & ...  & ...  &...     &11\\
Mrk 3         &  ...   & ...   & 2.90  & 3.77 &   86.0 & 109.0  & 207.0 &210.0 & ...   &82.0  & ...  &  ...   &1,16,16\\
Mrk 78        &  13.09 & 0.87  & 0.55  & 1.11 &   ...  &  ...   & ...   &  ... & ...   & ...  & ...  & ...    &7,1\\
Mrk 348       &  11.74 & 0.83  & 1.02  & 1.43 &   16.4 & 5.82   & 20.4  & 17.6 & ...   &12.2  & 9.81 &$<$0.083&20,27,14,14,19\\
Mrk 463E      &  ...  & ...    & 1.49  & 2.21 &   11.6 & 18.3   & 51.8  & 72.3 & ...   &13.5  & 30.3 &$<$0.008&27,17,17,19\\
Mrk 477       &  ...  & ...    & 0.51  & 1.31 &   ...  &   ...  & ...   &  ... & ...   &...   & ...  & ...    &29\\
Mrk 573       & 12.3  & 0.84   & 0.85  & 1.24 &$<$13.0&18.0     & 24.0  &  79.0& ...   &29.0  & ...  &...     &20,24,12,12\\
Mrk 1210      &  ...  & ...    & 2.08  & 1.89 &  ...   &  ...   & ...   & ...  & ...   &...   & ...  & ...    &1\\
NGC 424       &  ...  & ...    & 1.76  & 2.00 &   8.70 &  16.10 & 18.45 & 25.8 & $<$3.0&9.82  & 8.14 &$<$0.024&27,13,12,13,19\\
NGC 513       &  ...  & ...    & 0.48  & 0.41 &  12.76 &   1.91 & 4.43  &  6.54& 1.41  &14.50 & 27.49& 0.334  &27,13,13,13,19\\
NGC 591       &  9.77 & 1.12   & 0.45  & 1.99 &  ...   &  ...   & ...   &  ... & ...   &...   &...   & ...    &11,1\\
NGC 788       &  20.0 & 0.79   &$<$0.51& 0.51 &  ...   &  ...   & ...   & ...  & ...   & ...  & ...  & ...    &9,1\\
NGC 1068      &  13.14& 0.82   & 92.70 &198.0 &  700.0 & 970.0  &1600.0 &1900.0& 80.0  &550.0 &910.0 & ...    &20,27,12,12,12\\
NGC 2110      &  5.0  & 1.37   & 0.84  & 4.13 &  ...   &   ...  & ...   &  ... & ...   &...   &...   &...     &9,1\\
NGC 2273      &  5.77 & 0.86   & 1.36  & 6.41 &   ...  &  16.80 & 23.8  &  ... & ...   &...   & ...  & ...    &8,28,1\\
NGC 2992      &  ...  & ...    & 1.57  & 7.34 &  ...   &  ...   & ...   &  ... & ...   & ...  & ...  & 0.151  &30,19\\
NGC 3081      &  ...  & ...    & ...   & ...  &  ...   &  ...   & ...   &  ... & ...   &...   & ...  & ...    &\\
NGC 4388      &  11.15& 0.57   & 3.72  & 10.46& 76.6   &  46.1  & 106.0 &340.0 & ...   &85.1  &135.0 & 0.128  &8,27,14,14,19\\
NGC 4507      &  7.14 & 0.49   & 1.39  & 4.31 &   ...  &  18.40 & ...   &  ... & ...   &...   & ...  & ...    &9,1,1\\
NGC 5252      &  5.97 & 0.88   & ...   & ...  &  ...   &  ...   & ...   &  ... & ...   &...   & ...  & ...    &2\\
NGC 5506      &  8.91 & 0.95   & 4.24  & 8.44 &  59.0  &  26.0  & 58.0  & 135.0&  6.0  &81.0  &142.0 & 0.023  &10,27,12,12,12,19\\
NGC 5995      &  ...  & ...    & 1.45  & 4.09 & 16.50  &  6.13  & 8.47  & 12.90& ...   &5.32  & 25.40& 0.066  &27,14,14,19\\
NGC 6552      &  ...  & ...    & 1.17  & 2.57 &  ...   &  ...   & ...   & ...  & ...   &...   & ...  & ...    &27\\
NGC 7212      &  12.06& 0.69   & 0.77  &2.89  &  ...   &  ...   & ...   & ...  & ...   &...   & ...  & ...    &6,1\\
NGC 7314      &  ...  & ...    & 0.58  & 3.74 &  8.08  &  16.9  & 23.2  & 67.0 & ...   & 15.0 &14.2  &0.063   &1,14,14,19\\
NGC 7674      &  11.75& 0.89   & 1.79  & 5.64 &   18.0 &  31.0  & 46.0  & 46.0 & ...   &14.4  & 29.7 & 0.132  &11,27,16,14,19\\
NGC 7682      &  ...  & ...    & 0.22  & 0.47 &   ...  &  ...   & ...   &  ... & ...   &...   &...   &...     &27\\
Was 49b       &  ...  & ...    & ...   & ...  &  ...   &  ...   & ...   &  ... & ...   &...   & ...  & ...    &\\
\hline

\end{tabular}
{ \noindent \vglue 0.5cm {\sc Notes}: Column 1: Source name. Columns
2 and 3: Nuclear line ratios. Columns 4 and 5: Infrared flux (in
janskys) for 25 and $60~\mu m$. Columns $6-11$: Flux ($10^{-21}$W
$\rm cm^{-2}$) for \Neii~$12.81~\mu m$, \Nev~$14.32~\mu m$, \oiv~$
25.89~\mu m$, \Feii~$25.99~\mu m$, \Siii~$33.48~\mu m$, and \Si ii ~$
34.28~\mu m$, respectively. Column 12: Equivalent width in $\mu m$.
Column 13: References (for columns. 2-3, 4-5, 6-9, 10, 11-12 and 13,
respectively). The fluxes listed of reference 25 between the two
high-resolution modules, short-high and long-high, are not to be
directly compared because of the different aperture used
(Bernard-Salas et al. 2009).


$^a$ Uncertain detection for NGC 1068 and NGC 5506 (Sturm et al.
2002).

{\sc References}: (1) NED; (2) MPA-JHU DR7 ; (3) Brandl et al. 2006;
(4) Brightman \& Nandra 2008; (5) Armus et al. 1989; (6) Bennert et
al. 2006; (7) Contini et al, 1998; (8) Ho et al. 1997; (9) Vaceli et
al. 1997; (10) Kewley et al. 2000; (11) Veilleux \& Osterbrock 1987;
(12) Sturm et al. 2002; (13) Tommasin et al. 2008; (14) Tommasin et
al. 2010; (15) Veilleux et al, 2009; (16) Deo et al. 2007; (17)
Armus et al. 2004; (18) Weedman et al. 2005; (19) Wu et al, 2009;
(20) Sosa-Brito et al. 2001; (21) Dasyra et al. 2008; (22) Farrah et
al. 2007; (23) Dale et al. 2009; (24) Zhang \& Wang 2006; (25)
Bernard-Salas et al. 2009; (26) Goulding \& Alexander 2009; (27)
Tran 2003; (28) Sanders et al. 2003; (29) Imanishi. 2002; (30)
Surace et al. 2004;}
\end{table*}



\begin{table*}
\caption{\small The non-HBLR Sy2 sample}
\setlength{\tabcolsep}{2.5pt}
\renewcommand{\arraystretch}{1.2}

\begin{tabular}{lccccccccccccl} \hline \hline

name  & $\frac{[\rm O~III]\lambda5007}{H\beta}$ & $\frac{[\rm
N~II]\lambda6584}{H\alpha}$ & $f_{25}$ & $f_{60}$ & \Neii & \Nev&
\Neiii & \oiv & \Feii
& \Siii &  \Si ii &$ \rm EW$ &Reference \\
&  &  &  & &$12.81~\mu m $&$14.32~\mu m$&$15.56~\mu m$&$25.89~\mu
m$&$25.99~\mu m$&$33.48~\mu m$&$34.28~\mu m$&$6.2~\mu m$ \\
(1) & (2) &(3) & (4) & (5) & (6)&(7)& (8) & (9)& (10)& (11)&(12) &(13)&(14) \\
\hline

ESO 428-G014& ...  & ...  & 1.77 & 4.40 &  ...  &  82.90  & 168.01&  ...   & ...   &...   & ...   &...       &24,1\\
F00198-7926 & ...  & ...  & 1.15 & 3.10 & 6.19  & 12.27   & 14.03 & 33.03  & ...   & 17.13&...    &....      &27,13,13\\
F03362-1642 & ...  & ...  & 0.50 &1.06  & ...   &  ...    & ...   & ...    & ...   &...   & ...   &....      &24\\
F04103-2838 & ...  & ...  & 0.54 & 1.82 & 11.0  &  2.50   & 7.50  &   4.30 &$<$1.00&...   & ...   &....      &1,15,15\\
F04210+0401 & ...  & ...  & 0.25 & 0.60 &  ...  &  ...    & ...   &   ...  & ...   &...   & ...   &....      &24\\
F04229-2528 & ...  & ...  & 0.26 & 0.98 &  ...  &  ...    & ...   &   ...  & ...   &...   & ...   &....      &24\\
F04259-0440 & ...  & ...  & 1.41 & 4.13 & ...   &  ...    & ...   &   ...  & ...   & ...  & ...   &....      &1\\
F08277-0242 & ...  & ...  & 0.43 & 1.47 & ...   &  ...    & ...   &   ...  & ...   &...   & ...   &....      &24\\
F10340+0609 & ...  &...&$<$0.25  & 0.39 & ...   &  ...    & ...   &   ...  & ...   & ...  & ...   &....      &24\\
F13452-4155 & ...  & ...  & 0.81 & 1.84 & ...   &  ...    & ...   & ...    & ...   & ...  & ...   &...       &24\\
F19254-7245 & ...  & ...  & 1.35 & 5.24 & 31.48 &  2.77   & 13.19 & 6.35   & ...   & 9.07 & 56.80 &0.064     &27,22,22,19\\
F20210+1121 & ...  & ...  & 1.40 & 3.39 & ...   & ...     & ...   &   ...  & ...   &...   & ...   &...       &1\\
F23128-5919 & ...  & ...  & 1.59 & 10.80& 27.29 &  2.56   & 20.44 & 18.16  & ...   & 22.47& 17.48 &....      &24,22,22\\
IC 5298     & ...  & ...  & 1.80 & 9.76 & ...   &  ...    & ...   &  ...   & ...   &...   & ...   &....      &24\\
Mrk 334     & ...  & ...  & 1.05 & 4.35 & 30.0  & 13.0    & 26.0  &   15. 0& ...   &90.0  &...    &....      &1,16,16\\
Mrk 938     & 4.0  & 0.87 & 2.51 & 16.84& 52.1  &$<$2.19  & 6.37  &$<$0.66 & ... &$<$10.7 & 40.5  &0.44      &9,24,14,14,19\\
Mrk 1066    & 3.89 & 0.87 & 2.26 & 11.0 &  ...  & 17.80   & 52.0  & ...    & ...   &...   & ...   &...      &11,24,1\\
Mrk 1361    & 3.41 & 0.68 & 0.84 & 3.28 & ...   & ...     & ...   & ...    & ...   &...   & ...   &....      &2,24\\
NGC1143     & ...  &...   &$<$0.10&$<$1.10&  15.0&  ...    & ...   &  ...   & ...  & ...  & ...   &....      &30,1\\
NGC1144     & 2.94 & 0.84 & 0.62 & 5.35 & 17.2  & 0.92    & 5.40  &  5.31  & 2.51 &32.2  &67.0   &0.343      &9,24,14,14,14,19\\
NGC1241     &5.56  & 0.92 & 0.60 & 4.37 &  13.4 &  1.61   & 8.08  & 5.46   & 2.41 & 13.8 & 19.1  &0.461      &9,24,14,14,14,19\\
NGC1320     & ...  & ...  & 1.32 & 2.21 & 9.0   & 8.0     & 9.0   &  32. 0 & ...  &12.0  &10.4   &0.082      &24,16,14,19\\
NGC1358     & 7.69 &1.89 &$<$0.12& 0.38 &  ...  &  ...    & ...   &  ...   & ...  &...   &...    &....       &9,24\\
NGC1386     & 16.67& 1.94 & 1.46 & 6.01 & 17.8  & 34.5    & 36.6  & 106.0  & ...  &31.0  & 32.8  &0.053      &9,24,14,14,19\\
NGC1667     & 7.58 & 2.38 & 0.67 & 6.29 &  10.1 & 1.32    & 7.23  &   7.06 & ...  &24.5  & 54.3  &0.391      &8,24,14,14,19\\
NGC1685     & ...  & ...  & 0.22 & 0.98 & ...   &  ...    & ...   &  ...   & ...  &...   &...    &....       &24\\
NGC3079     &  ... &  ... & 3.65 & 50.95& 98.96 &   1.31  & 23.17 &   8.45 & 14.57 &56.88 &173.51&0.458      &24,25,25,25,19\\
NGC3147     & 6.11 & 2.71 & 1.08 & 8.40 & ...   & ...     & ...   &  ...   & ...   &...   &...   & ...       &4,8\\
NGC3281     & 10.0 & 0.98 & 2.63 & 6.73 & ...   & ...     & ...   & ...    & ...   & ...  & ...  &...        &10,28\\
NGC3362     & 8.14 & 1.74 & 0.35 & 2.13 &  ...  & ...     & ...   &  ...   & ...   &...   &...   &....       &2,24\\
NGC3393     & 10.2 & 1.2  & 0.75 & 2.25 & ...   &   42.40 & 95.0  &  ...   & ...   &...   & ...  &...        &20,24,1\\
NGC3660     & 2.63 & 0.82 & 0.64 & 2.03 &  6.51 &    0.98 & 1.49  &  3.61  & ...   &9.52  &9.54&$<$0.434     &4,27,13,13,19\\
NGC3982     & 11.97& 0.94 & 0.97 & 7.21 &   11.4&  2.89   & 6.79  & 5.11   & ...   &15.4  & 32.8 &0.467      &2,24,14,14,19\\
NGC4117     & 3.98 &  0.41& ...  &  ... & ...   &  ...    & ...   &  ...   & ...   &...   &...   &....       &2\\
NGC4501     & 5.25 & 2.1  & 3.02 &19.93 &  7.02 &$<$1.5   & 4.72  &   4.22 & 2.62  &7.9   &16.70&$<$0.187    &4,27,13,13,13,19\\
NGC4698     & 4.29 &1.31&$<$0.154&0.26  &  ...  &  ...    & ...   &   ...  & ...   & ...  & ...  &...        &8,1\\
NGC4941     & ...  & ...  & 0.46 & 1.87 &  13.50&   8.21  & 24.8  &   32.8 & ...   &5.79  &13.4&$<$0.041     &24,14,14,19\\
NGC5128     & ...  & ...  & 28.2 &213.0 &  221.0&   27.0  & 141.0 &  98.0  & 12.0  & 223.0&545.0 &0.014      &24,12,12,12,18\\
NGC5135     & 3.57 & 0.85 & 2.39 &16.60 & 36.7  &   4.88  & 16.7  &  71.3  & 6.8   &38.30 & 140.0&0.384      &9,24,14,14,14,19\\
NGC5194     & 8.96 & 2.90 & 17.47&108.7 &   ... &...      & ...   &   ...  & ...   &...   & ...  & 0.372     &8,8,19\\
NGC5256     & 4.47 & 0.62 & 1.07 & 7.25 &  57.04&   7.96  & 27.95 &   52.94& 4.35  &50.93 & 87.02&0.608      &11,28,25,25,25,19\\
NGC5283     & ...  &  ... & 0.09 & 0.13 & ...   &   ...   & ...   &  ...   & ...   &...   & ...  &...        &1\\
NGC5347     & ...  & ...  & 0.96 & 1.42 & 4.17  &   2.08  &4.09   &  7.64  & ...   &$<$3.38&$<$4.62&$<$0.112 &24,14,14,19\\
NGC5643     & 11.4 & 1.04 & 3.65 & 19.50&  46.41&  24.63  & 56.47 & 118.28 & ...   &58.0  & 55.0 & ...       &6,24,26,12\\
NGC5695     & 10.47& 1.48 & 0.13 & 0.57 &  ...  &  ...    & ...   &  ...   & ...   &...   & ...  &....       &11,24\\
NGC5728     & 11.8 & 1.4  & 0.88 & 8.16 &   ... &  ...    & ...   & ...    & ...   &...   & ...  &...        &20,24\\
NGC5929     & 3.42 & 0.59 & 1.67 & 9.52 &  13.2 &  1.14   & 9.83  &  5.32  & ...   &6.45 &21.5&$<$0.48       &2,27,14,14,19\\
NGC6300     & 9.09 & 2.23 & 2.27 & 14.70&  11.52&  12.54  & 15.28 &  29.45 &$<$3.49&$<$7.63&11.31& ...       &9,24,26,26,26\\
NGC6890     & ...  & ...  & 0.65 & 3.85 &  11.32&    5.77 & 6.57  &  10.10 & ...   & 16.97&26.54&0.237       &24,13,13,19\\
NGC7130     & ...  & ...  & 2.16 & 16.71&  79.3 &  9.09   & 29.4  &   19.7 & 11.81 &48.2  & 93.9&0.493       &28,14,14,14,19\\
NGC7172     & 10.0 & 1.0  & 0.95 & 5.74 &  33.0 &  10.2   & 17.1  &  45.4  & ...   &26.9  & 59.3&0.045       &9,24,14,14,19\\
NGC7496     & 0.34 & 0.46 & 1.93 & 10.14& 48.08 & $<$1.8  & 6.67  &$<$2.4  & ...   &39.47 &44.58&0.912       &10,28,13,13,19\\
NGC7582     & 2.33 & 0.69 & 7.48 & 52.47& 148.0 & 22.0    & 67.0  &  116.0 & 8.0   &113.0 &218.0&0.274       &20,24,12,12,12,19\\
NGC7590     & 5.00 & 1.05 & 0.89 & 7.69 &  7.78 & $<$1.5  & 3.49  & 5.60   & ...   &15.69 &26.50&0.496       &9,28,13,13,19\\
NGC7672     &  ... & ...  &$<$0.15&0.46 & ...   &   ...   & ...   &   ...  & ...   &...   &...  &....        &24\\
UGC6100     & 12.63& 1.11 & 0.20 & 0.57 &  ...  &  ...    & ...   &  ...   & ...   &...   &...  &....        &2,1\\
\hline

\end{tabular}

{\noindent \vglue 0.5cm {\sc Note}: This table has the same Columns
and References as the table 2.}


\end{table*}





\def\oiii{\ifmmode [O {\sc iii}] \else [O {\sc iii}]\ \fi}
\def\Neii{\ifmmode [O {\sc ii}] \else [Ne {\sc ii}]\ \fi}
\def\Neiii{\ifmmode [O {\sc iii}] \else [Ne {\sc iii}]\ \fi}
\def\Nev{\ifmmode [O {\sc v}] \else [Ne {\sc v}]\ \fi}
\def\oiv{\ifmmode [O {\sc iv}] \else [O {\sc iv}]\ \fi}
\def\Feii{\ifmmode [Fe {\sc ii}] \else [Fe {\sc ii}]\ \fi}
\def\Siii{\ifmmode [S {\sc iii}] \else [S {\sc iii}]\ \fi}
\def\Siiv{\ifmmode [O {\sc ii}] \else [Si {\sc ii}]\ \fi}




\begin{table*}
\caption{\small The Starbursts Galaxies sample}
\setlength{\tabcolsep}{3.5pt}
\renewcommand{\arraystretch}{1.2}
\begin{tabular}{lccccccccccccl} \hline \hline

name  & $\frac{[\rm O~III]\lambda5007}{H\beta}$ & $\frac{[\rm
N~II]\lambda6584}{H\alpha}$ & $f_{25}$ & $f_{60}$ & \Neii & \Nev
&\Neiii & \oiv &  \Feii
& \Siii & \Si ii &$ \rm EW$ &Reference \\
&  &  &  & &$12.81~\mu m $&$14.32~\mu m$&$15.56~\mu m$&$25.89~\mu
m$&$25.99~\mu m$&$33.48~\mu m$&$34.28~\mu m$&$6.2~\mu m$ \\
      (1) & (2) &(3) & (4) & (5) & (6)&(7)& (8) & (9)& (10)& (11)&(12) &(13)& (14) \\
\hline

NGC253      & 0.36 & 0.72 & 119.7& 784.2&2832.33&$<$20.5  &204.64 & 154.74 &250.56 &1538.03&2412.03&...   &10,1,25,25,25     \\
NGC520      & 0.74 & 0.39 &  3.22& 31.52& 44.62 &$<$0.64  & 7.53  &  8.1   & 7.56  & 89.44 &190.71 &0.563 & 8,3,25,25,25,3   \\
NGC660      & 2.53 & 0.85 &  7.30&65.52 &353.01 &$<$2.26  & 36.96 & 18.8   & 22.92 &246.07 &441.96 & 0.504& 8,3,25,25,25,3   \\
NGC1097     & 5.10 &  3.0 &  7.30&53.35 &37.24  &$<$0.43  &  6.33 & 4.41   & 11.48 &100.05 &225.19 &0.459 & 20,3,25,25,25,3  \\
NGC1222     & 2.29 &  0.21& 2.28 & 13.06&80.57  &$<$0.57  & 89.41 & 9.92   & 4.74  &132.52 &112.05 & 0.624& 10,3,25,25,25,3  \\
NGC1365     &  3.0 &  0.55& 14.28&94.31 &139.50 &18.83    & 59.53 &141.57  & 22.2  &246.89 &500.3  &0.368 & 20,3,25,25,25,19 \\
IC342       & ...  & ...  & 34.48& 180.8&615.46 &$<$2.41  & 37.2  &$<$7.7  & 51.87 &672.46 &985.73 &0.497 & 3,25,25,25,3     \\
NGC1614     & 0.81 & 0.60 &  7.50& 32.12& 249.0 &$<$0.99  &63.32  & 8.68   & 12.99 & 101.06& 148.6 &0.561 & 10,3,25,25,25,3  \\
NGC2146     &  0.48& 0.45 & 18.81&146.69& 625.0 &$<$2.81  &91.16  & 19.33  & 52.69 &848.02 &1209.35&0.545 & 8,3,25,25,25,3   \\
NGC2623     & ...  & ...  & 1.81 &23.74 & 55.55 & 2.71    & 15.08 & 9.62   & 2.23  & 13.78 & 28.74 &0.598 & 3,25,25,25,3     \\
NGC3256     &  ... &   ...& 15.69&102.63&514.19 &$<$2.35  & 64.42 & 12.23  & 33.56 & 484.64&623.37 &0.603 & 3,25,25,25,3     \\
NGC3310     &  0.95& 0.66 & 5.32 &34.56 & 27.57 &$<$0.25  & 28.35 & 4.03   & 8.78  &106.02 & 143.54&0.789 &8,3,25,25,25,3    \\
NGC3556     & 0.26 & 0.32 &  4.19&32.55 & 21.47 &$<$0.37  & 3.23  &$<$1.50 & 3.14  & 96.37 & 96.44 & 0.502&8,3,25,25,25,3    \\
NGC3628     & 1.77 & 0.95 &  4.85&54.80 &125.63 & 0.9     & 10.05 &$<$2.37 & 11.96 &149.93 & 277.55& 0.50 &8,3,25,25,25,3    \\
NGC4088     & 0.21 &  0.32&  3.45& 26.77& 37.03 &$<$0.39  & 2.45  & 0.73   &  2.15 & 35.02 & 63.33 & 0.496&8,3,25,25,25,3    \\
NGC4194   &$\le$1.35&0.46 &  4.51& 23.20&165.47 & 2.96    & 53.99 & 27.45  & 8.48  &151.01 &185.82 & 0.529&5,3,25,25,25,3    \\
Mrk52       &  ... &   ...& 1.05 & 4.73 & 29.38 &$<$0.57  & 3.82  & 1.31   & 1.90  & 60.43 & 38.02 & 0.535& 3,25,25,25,3     \\
NGC4676     & ...  & ...  & 0.33 & 2.67 & 27.78 &$<$0.25  & 4.68  & 1.37   &  1.78 & 23.95 & 41.27 & 0.61 & 3,25,25,25,3     \\
NGC4818     & 0.15 & 0.63 &  4.40& 20.12&184.93 &$<$1.21  & 13.73 &$<$1.59 & 10.83 & 80.54 & 128.66& 0.459&10,3,25,25,25,3   \\
NGC4945     &   ...&   ...& 42.34&625.46&583.84 & 3.84    & 68.98 & 47.85  & 55.21 &338.91 & 846.23& 0.432& 3,25,25,25,3     \\
NGC7252     & ...  &...   &  0.43& 3.98 & 41.68 &$<$0.33  & 3.72  &  1.33  & 3.12  & 24.75 & 70.53 & 0.585& 3,25,25,25,3     \\
NGC7714     &  1.35&  0.35& 2.88 & 11.16& 102.55&$<$1.0   & 77.42 & 5.51   & 5.93  & 115.9 &102.97 & 0.601&10,3,25,25,25,3   \\
\hline
\end{tabular}

{\noindent \vglue 0.5cm {\sc Note}: This table has the same Columns
and References as the table 2.}

\end{table*}


\section{Sample and Data Reductions}\label{S:sample}

Based on the related analysis between HBLR and non-HBLR Sy2s, Wu et
al. (2011) suggested that the former and follower are dominated by
AGNs and starburts, respectively. In Section 1, many works showed
that Seyfert galaxies could experience some evolution processes. To
investigate the evolutionary probability, we define a sample
composed of Seyfert and starburst galaxies. However, due to both
considering sample completeness and having as large a sample size as
possible, this is not an easy job. There are two criteria on sample
selection mainly arising from: (1) the spectropolarimetric
observations are determinant to distinguish between HBLR and
non-HBLR Sy2s; (2) all sources of our multiple subsamples have
within $z\sim$0.18. Next we introduce the sample selection of each
of our multiple subsamples in detail.

Starburst subsample derives from the study object of Bernard-Salas
et al. (2009) which is well known nearby starburst
galaxies. Such starburst system often depends on high nuclear star
formation rates as the primary energy source. It now seems clear that producing the most luminous infrared
galaxies is trigged by strong interactions and mergers of molecular
gas-rich spirals and that an important link between starburst galaxies
and the AGN phenomena is very likely represented by luminous
infrared galaxies (LIRGs; $\geq10^{11}L_{\sun}$) (Sanders $\&$
Mirabel 1996).

Considering that Mrk 266 and NGC 3079 are suggested by Tran (2003)
as non-HBLR Sy2s, so our starburst subsample has 22 objects. Five
out of the seven objects with a known AGN component show emission of
the high excitation \NeV~line, meanwhile 15 objects are not signs of
AGN activity. In addition, seven of this subsample are LIRGs
(Bernard-Salas et al. 2009). In the most luminous of these galaxies,
Lagache et al. (2005) suggested that AGN activity is very common,
even though their energy output is not dominated by the activity.
Based on the above analysis, we have some criteria on sample
selection mainly arising from: (1) only those sources have within
$z<0.03$; (2) the \NeV/\NeII~ ratio of $\sim$0.13 is the upper
limit.

To avoid the disadvantages of sample misclassification, we try to
collect updated and more precise classifications from the
literatures. So we mainly employ the combined CfA and 12 $\mu$m
sample of Tran (2003) as our Sy1 subsample. Tran (2003) had revised
the original Sy1 table of Rush et al. (1993) to include only Sy1 and
Sy1.5 galaxies, excluding those classified as Sy1.8 or Sy1.9. In
order to avoid biases, he had also removed several highly
radio-luminous 3C galaxies from their Sy1 sample, as noted in their
table 2. In addition, we exclude NGC 2992 from the Sy1 sample,
because the object is classified by NASA/IPAC Extragalactic Database
(NED) as a Sy2 galaxy and the spectropolarimetric observation shows
that it is an HBLR galaxy (Lumsden et al. 2004).

The Sy1 subsample comprises a subset of the combined 12 $\mu$m and
CfA samples of AGNs. To eliminate some forms of selection effects or
incompleteness, Rush et al. (1993) followed the approach originated
by Spinoglio $\&$ Malkan (1989), a selection based on a flux limit
at 12 $\mu$m which minimizes wavelength-dependent selection effects.
Using an IRAS detection at 12$\mu$m, $F_{12}(IRAS) >0.22$~Jy, and color
selection $F_{60}(IRAS)> 0.5F_{12}(IRAS)$ or $F_{100}(IRAS) >
F_{12}(IRAS)$ defines the parent sample to remove stars but few
galaxies (Rush et al. 1993; Gallimore et al. 2010). The CfA sample
had also been defined by Osterbrock $\&$ Martel (1993). So our
sample selection mainly meet the following criteria: (1) only those
sources have within $z<0.11$; (2) this subsample includes only Sy1
and Sy1.5 galaxies.

Compared to the classical Seyfert 1 galaxies and starburst galaxies,
only a small percentage of Sy2s have been observed by
spectropolarimetry, and they are not extensively well studied,
because this observation is a time-consuming and tedious job (Tran
2001; Wang $\&$ Zhang 2007). Since it is difficult in the
spectropolarimetric observations for both HBLR and non-HBLR Sy2s, we
have constructed an amalgamated sample by searching all the
literatures related to Seyfert galaxy samples.

Some previous samples for studying HBLR and non-HBLR Sy2s are the
amalgamation of different observations with diverse quality of
spectropolarimetric data (e.g., Gu $\&$ Huang 2002; Zhang $\&$ Wang
2006; Shu et al. 2007), varying from object to object determined by
the brightness, observers, integration time, and other factors. For
example, although Ruiz et al. (1994) had found that it was no sign
for broad H$\alpha$ and H$\beta$ components in the polarized light
for Mrk 334, its flux was regarded as polarized broad-line derives
from a maximum of $\sim0.6\%$ of the light in H$\alpha$. For other
Sy2s with detected PBLs, the mean percentage of $1\%-5\%$ has been
reported. Obviously, it is necessary to evaluate the sensitivities
of detecting PBLs in different groups from which our data are taken
(Gu $\&$ Huang 2002).

The situation we are faced with has some changes which some groups
(e.g., Young et al. 1996; Lumsden et al. 2004; Shi et al. 2010; Tran
et al. 2010) have presented detailed information for their polarized
broad emission components, so that we can derive their sensitivities
--- the detected mean polarized broad H$\alpha$ flux is about
$2\times10^{-15}$ergs s$^{-1}$ for NGC 3147 and NGC 4698, and their
mean percentages of polarized broad H$\alpha$ emission to total
H$\alpha$ emission is $0.11\%\pm0.01\%$ and $0.31\%\pm0.01\%$,
respectively (Tran et al. 2010). However, most sources of our sample
have not such comparable sensitivities.

The significant case is the difference in judging either non-HBLR or
HBLR Sy2s of two Sy2s (NGC 5347 and NGC 5929). The two sources show
faint polarized broad emission lines with high signal-to-noise ratio
(S/N) data discovered by Moran et al. (2001), and they were regarded
as non-HBLR Sy2s in Tran's (2001) sample (Gu $\&$ Huang 2002).
However, NGC 5929 was also reported as a non-HBLR Sy2 galaxy by
Lumsden et al. (2001). Another case is the classification of Mrk
573 which was suggested as a non-HBLR Sy2 galaxy (Tran 2001; 2003),
and now it is regarded as an HBLR Sy2 galaxy (Nagao et al. 2004; Ramos
Almeida et al. 2008).

Based on the above analysis, the subsamples' selection of HBLR and
non-HBLR Sy2s mainly meet the following criteria: (1) all sources of
the two subsamples have within $z\sim$0.18; (2) the percentage of
the polarization is $>1\%$ for HBLR Sy2s and $<1\%$ for non-HBLR
Sy2s. With regard to their spectropolarimetric observations of HBLR
and non-HBLR Sy2s in our sample, Wu et al. (2011) described them
(here, Mrk 573 is regarded as an HBLR Sy2 galaxy) in their Table A1 in
detail.

Since Seyfert 1 and Seyfert 2 subsamples are included in our sample,
we discuss how it would be possible to disentangle evolutionary
effects from geometric effects. Baum et al. (2010) found that in
Seyfert sample, the opacity of the torus at \NeV~14.32 $\mu$m is not
large, so we suggest that the \NeV~14.32 $\mu$m emission is roughly
independent orientation. In addition, \oiv~25.89 $\mu$m originates
purely from the narrow-line region (NLR) in AGNs. In a unified
scheme, the NLR line luminosity should be independently orientated
(Sturm et al. 2002). Because \NeV~14.32 $\mu$m and \oiv~25.89 $\mu$m
are mainly used as the diagnostic of AGNs, we may remove basically
their geometric effects in our sample.

To avoid sample bias due to the absence of spectropolarimetric
observations, we exclude 18 objects, 4 HBLR Sy2s and 14 non-HBLR
Sy2s, from Table 5 in Wang $\&$ Zhang (2007). We construct a sample
composed of 170 Seyfert galaxies and starburst galaxies, which
contain 45 Sy1s, 46 HBLR Sy2s, 57 non-HBLR Sy2s, and 22 starburst
galaxies. The subsample of Sy1s for the combined CfA and 12 $\mu m$
sample is shown in Table 1. The subsamples of HBLR and non-HBLR Sy2s
are the amalgamation of different observations with diverse quality
of spectropolarimetric data, and they are listed in Tables 2 and 3,
respectively. Starburst subsample derives from the study object of
Bernard-Salas et al. (2009) and is presented in Table 4.

 To improve the accuracy of the measurements of our sample
objects, we use the catalog of MPA-JHU emission line measurements
for the Sloan Digital Sky Survey (SDSS) Data Release (DR7). The
measurements are available for ~927552~ different spectra. Compared
to previous DR4 release, this represents a significant extension in
size and a number of improvements in the data. These data are
available online at the following address: http://www. mpa-garching.
mpg. de/SDSS/DR7/. Each fiber is 3\arcsec~ in diameter (Bernardi et al.
2003; Adelman-McCarthy et al. 2006). We do cross-correlation with
the NED and exclude those objects which are larger than 2\arcsec~ in RA
or DEC errors. According to S/N $>5$ for
the H\b~\l 4861, H\a~\l 6563, \NII~\l 6584, and \oiii~\l 5007 lines, meanwhile
 we exclude the objects with low emission-line signal-to-noise ratio
  (S/N).

In order to derive reliable statistical results, we have the two
highest priorities obtaining a sample which is complete and
larger. Since data collected from the literatures or datasets usually
involve their different quality, they can also influence the
results. In this paper, although they may have some selection effects,
we use the line ratios among various fluxes to study Seyfert and
starburst galaxies, so the selection effects could little or not
influence our results.

\section{Derived Quantities}\label{S:Quantities}

Baldwin, Phillips, \& Terlevich (1981) first proposed using optical
emission-line ratios to classify the dominant energy source in
emission line galaxies. Later, Kewley et al. (2001a) developed the first purely theoretical
classification scheme to derive a ``starburst line" on the BPT
diagrams. However, they had not obtained the accurate line between
star forming galaxies and AGNs. To obtain a more stringent sample of
star-forming galaxies, Kauffmann et al. (2003b) shifted the Kew01
line to form, Ka03 line, a semi-empirical upper boundary for the
star-forming branch observed with the SDSS.

The \oiii \l 5007/H\b~versus \NII~\l 6584/H\a~diagram is most
commonly used to separate starburst galaxies from AGN hosts (see
e.g., Brinchmann et al. 2004; Lamareille et al. 2004; Mouhcine et
al. 2005; Gu et al. 2006). Because dust is often heavily enshrouded
nuclei, an important limitation of galaxy optical diagnostic is the
effect of extinction. The advent of sensitive infrared line data
from the $Infrared~ Space~ Observatory ~(ISO)$ was vital to gazing
deeply into buried nuclear sources (Genzel et al. 1998; Laurent et
al. 2000a; Sturm et al. 2002; Peeters et al. 2004). Based on the
strength of PAH emission features correlating with mid-infrared line
ratios, ionization-sensitive indices are first showed by Genzel and
collaborators. \spitzer~ Space Telescope affords the sensitivity and
wide bandpass, so the mid-IR mechanism has a wide prospect (Ho
2008).

The \spitzer~ with unprecedented sensitivity and angular resolution
can achieve a more detailed view to investigate the nature of
galactic nuclei (e.g., Armus et al. 2004; Smith et al. 2004). In
Figure 4, we compare the values of \Nev~14.32 $\mu m$/\Neii~12.81
$\mu m$ with \oiv~25.89 $\mu m$/\Siii~33.48 $\mu m$ measured in our
sample of Seyfert galaxies and starburst galaxies. In Figure 5,
we investigate the positions of galaxies on the \Feii~
25.99 $\mu m$/\oiv~25.89 $\mu m$ versus \oiv~25.89 $\mu m$/\Siii~
33.48 $\mu m$ diagnostic diagram. These diagrams separate starburst
galaxies from those dominated by AGNs. In Figure 6, since the
$f_{60}/f_{25}$ ratio denotes the relative strength of starburst and
AGN emissions (Wu et al. 2011), we show the correlation of the
$f_{60}/f_{25}$ ratio versus 6.2 $\mu $m EW.

\begin{figure}
\centerline{\includegraphics[scale=0.5,angle=0]{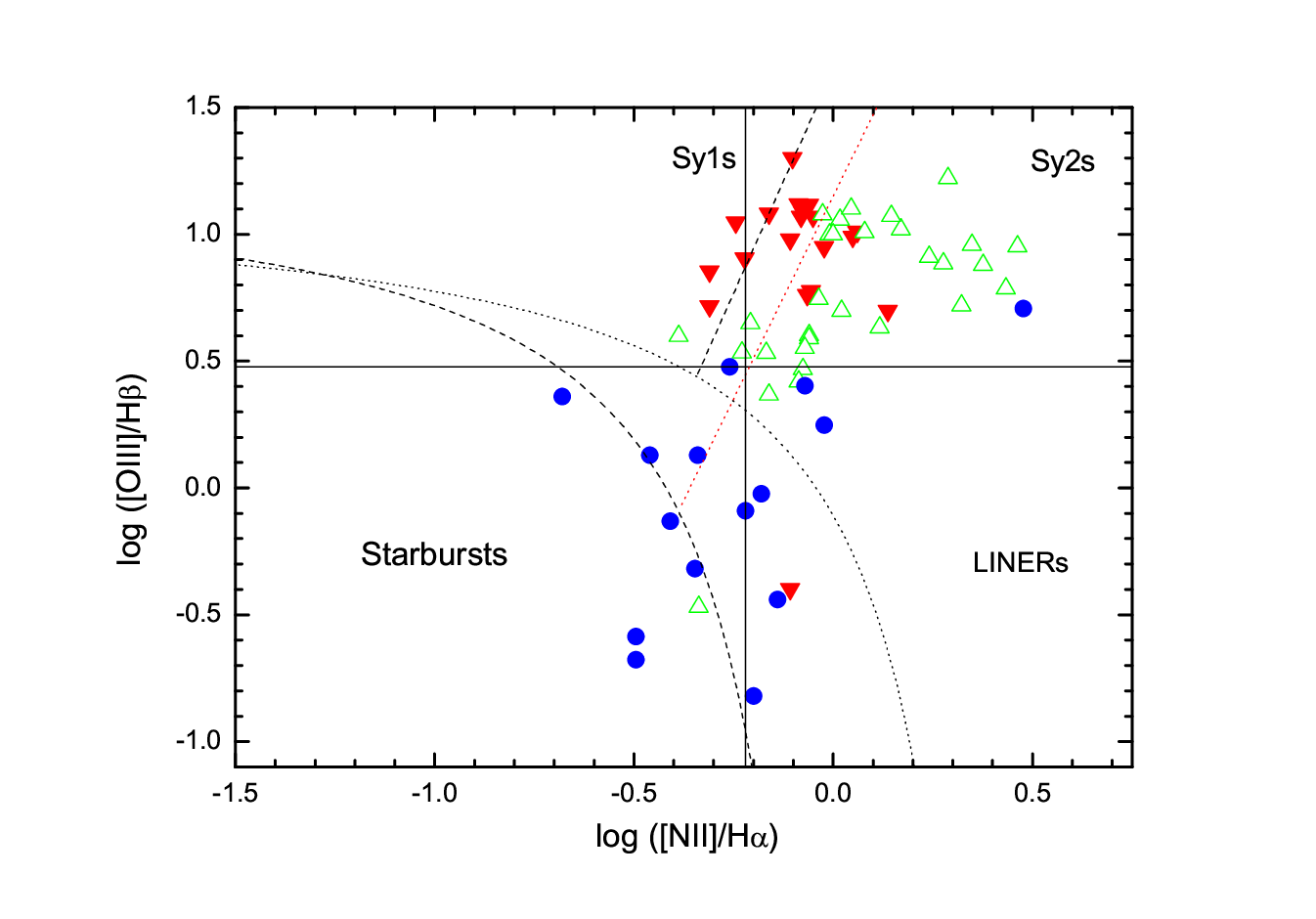}}
\caption{Traditional diagnostic diagram. Red filled and green open
triangles mark HBLR Sy2s and non-HBLR Sy2s; blue filled circles mark
starbursts. The dotted curve is the theoretical ``maximum starburst
line" derived by Kewley et al. (2001a) as an upper limit for
star-forming galaxies; the dashed curve on the diagram is the
Kauffmann et al.' (2003a) semi-empirical lower boundary for the
star-forming galaxies; the dashed line is Zhang et al. (2008); the
red dotted line is described in the text. [See the electronic
edition of the Journal for a color version of this figure.]}
\end{figure}

\section{Results}\label{S:types}

In Section 4.1, we show the separations of starburst galaxies,
non-HBLR Sy2s, and HBLR Sy2s with optical diagnostic diagrams. In
Section 4.2, their separations are presented by infrared spectral
diagnostics and are confirmed by statistics. In Section 4.3, we
demonstrate their possible evolution with the change of ~\NII~ flux
between starburst galaxies, non-HBLR Sy2s, and HBLR Sy2s/Sy1s.

\subsection{Optical Diagnostic Diagrams}\label{S:types}

\begin{figure*}
\centerline{\includegraphics[scale=0.65,angle=0]{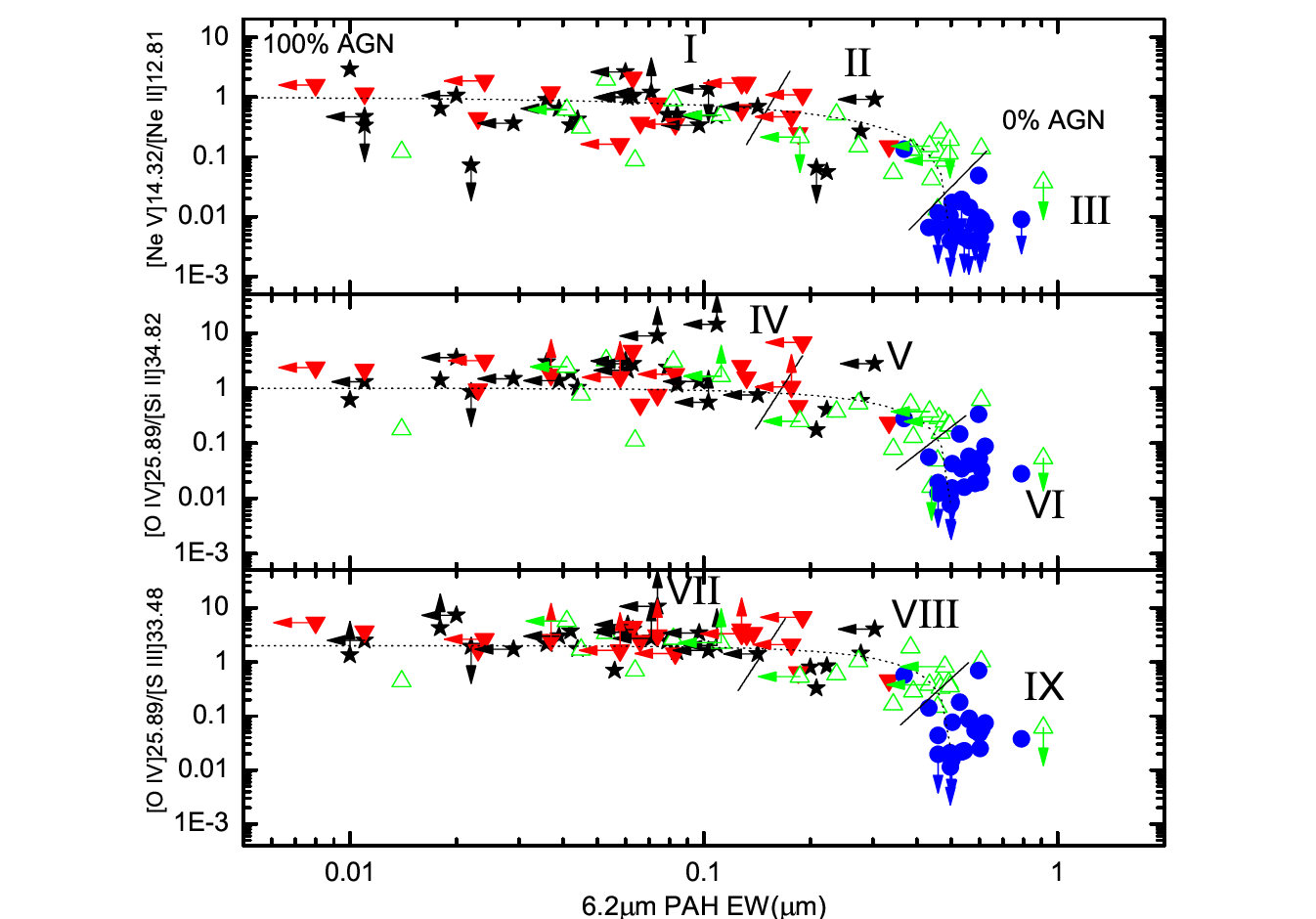}}
\caption{Ratios of mid-infrared forbidden lines as a function of the
$6.2~\mu m$ PAH feature equivalent width. Red filled and green open
triangles mark HBLR Sy2s and non-HBLR Sy2s, respectively; blue
filled circles mark starbursts; black stars mark Sy1s. The dotted
lines are linear mixing models of a ``pure" AGN and a ``pure"
star-forming source (see text). The solid lines and Roman numerals
delineate regions distinguished by Sy1s, HBLR Sy2s, non-HBLR Sy2s,
and starbursts (see table 5). [See the electronic edition of the
Journal for a color version of this figure.]}
\end{figure*}

An excellent means of classifying galaxies by easily measured line
ratios is provided by the BPT diagnostic diagrams. Here we employ
the line ratio of \oiii~ \l 5007/H\b~ against the \NII~ \l 6584/H\a~
ratio. Starburst galaxies fall onto the lower left-hand region of
these plots, narrow-line Seyferts are located in the upper right and
LINERs are in the lower right-hand zone (Kewley et al. 2001b). We
present the diagram for the galaxies of our subsamples in Figure 1
(due to the enough large size of the Sy1s sample of Zhang et al.
2008, we do not put our Sy1s sample into Figure 1). Equations 1 and
2 display the Kewley et al.'s (2001a) theoretical curve (the dotted curve
in Figure 1) and Kauffmann et al.'s (2003) semi-empirical curve (the
dashed curve), respectively.

\begin{equation}
\rm log (\frac{[O III]{5007}}{H\beta})= \frac{0.61}{log ([N II]/
H\alpha)-0.47}+1.19,
\end{equation}

\begin{equation}
\rm log (\frac{[O III]{5007}}{H\beta})= \frac{0.61}{log ([N II]/
H\alpha)-0.05}+1.3,
\end{equation}

\begin{equation}
\rm log (\frac{[O III]{5007}}{H\beta})= 3.53\times log ([N II]/
H\alpha)+1.65,
\end{equation}

Based on the galaxy catalog of SDSS Data Release 4 (DR4) at redshift
$z < 0.3$, Zhang et al. (2008) found that Seyfert 1 and Seyfert 2
galaxies have different distributions on the \oiii ~\l 5007/H\b~
versus \NII~ \l 6584/H\a~ diagram. Seyfert 2 galaxies display a
clear left boundary on the BPT diagram, while Seyfert 1 galaxies do
not show such a cutoff. They defined a line (dashed line on this
diagram) with the equation 3.

\begin{equation}
\rm log (\frac{[O III]{5007}}{H\beta})= 3.2\times log ([N II]/
H\alpha)+1.15,
\end{equation}

\begin{table*}
\caption{Classifications by Region in Figure 2}
\begin{small}
\begin{center}
\begin{tabular}{lccccl}
\hline \hline

Region  & Number of Sources & Sy1s (\%) & HBLR Sy2s (\%) & non-HBLR
Sy2s (\%) & Starbursts (\%) \\
 (1)& (2)& (3)& (4)& (5)& (6)\\
\hline
I     &  41   &   51  &  32   & 17     &  0  \\
II    &  23   &   17.5&  17.5 & 61     &  4  \\
III   &  23   &    0  &  0    &  9     & 91  \\
IV    &  40   &   50  &  33   & 17     &  0  \\
V     &  21   &   19  &  14   & 62     &  5  \\
VI    &  24   &    0  &   0   & 12     & 88  \\
VII   &  43   &   51  &  33   & 16     &  0  \\
VIII  &  20   &   25  &  20   & 50     &  5  \\
IX    &  26   &    0  &   0   & 19     & 81  \\
\hline

\end{tabular}
\parbox{6.5in}
{\baselineskip 9pt \noindent \vglue 0.5cm {\sc Note}: Column 1: the
regions in Figure. 2. Column 2: number of Sources. Column 3: the
percents in Sy1s. Column 4: the percents in HBLR Sy2s. Column 5: the
percents in non-HBLR Sy2s. Column 6: the percents in starburst
galaxies.}
\end{center}
\end{small}
\end{table*}


The non-HBLR Sy2s have weaker optical narrow-line ratios than the
HBLR Sy2s on the BPT diagram, so they are closer to the sequence of
star-forming galaxies on the BPT diagram (Deo et al. 2009), which
appears consistent with that non-HBLR Sy2s are dominated by
starbursts (Wu et al. 2011).

In Figure 1, we define the red dotted line with the equation 4 as
the separation of HBLR and non-HBLR Sy2s. The line can clearly
distinguish between the two types of objects: HBLR Sy2s mainly
display a left boundary (12/19) while non HBLR Sy2s mostly display a
right boundary (26/30). Compared to non-HBLR Sy2s, HBLR Sy2s are
closer to Sy1s in the diagram. Since our sample is not enough
completeness, this boundary may not quite accurate and may have no
strict physical significance.

\subsection{Infrared Spectral Diagnostics of Nuclei and Starburst Regions}\label{S:types}

\subsubsection{Emission-Line Ratios and PAH Strength}\label{S:types}

Next we use the different emission-line ratios as the mid-infrared
diagnostic. The top panel of Figure 2 uses the 6.2 $\mu m$ PAH
feature and the emission-line ratio \NeV~14.32 $\mu m$/\Neii~12.81
$\mu m$ in a similar mid-infrared diagnostic diagram. Due to the
intense star formation in the nuclear region of many active
galaxies, some fraction of the measured fluxes of low lying fine
structure line (excitation potential $\le$50 eV) will be produced by
photoionization from stars rather than AGNs, while the high
excitation lines (\NeV, \oiv) show little or no contamination from
possible starburst components (Sturm et al. 2002). Since the
ionization potentials needed to product $\rm Ne^{4+}$ and
 $\rm Ne^+$ are 97.1 and 21.6 eV, respectively, the above result
 can be understood. Starburst galaxies and non-HBLR Sy2s exhibit comparatively large
$6.2~\mu m$ EWs and relatively low ratios of \NeV~14.32~$\mu
m$/\Neii~$12.81~\mu m$, predicating that the strong contribution
comes from \Neii~$12.81~\mu m$ and is negligible from AGN emission.

We also use another two diagnostic diagrams in Figure 2.
High-ionization lines like \oiv~25.89 $\mu m$ are somewhat difficult
to be detected for its higher ionization potential. Removing an
electron from doubly ionized oxygen is required at least $54.9$ eV.
Both the middle and bottom panels involve the more easily detectable
\Si ii~34.82 $\mu m$ line which has an ionization potential of 8.15
eV; and another strong mid-infrared lines involved is \Siii~
$33.48~\mu m$ with the ionization potentials of 23.3 eV (Dale et al.
2006). In the middle and bottom panels, starburst galaxies and
non-HBLR Sy2s exhibit comparatively large $6.2~\mu m$ EWs. Starburst
galaxies have rough trends in the three panels,
which significantly exhibit the comparatively small ratios,
so the difference in the ionization potential between various
emission lines may be the main factor.

\begin{table*}
\caption{Classification by Region in Figure 3}
\begin{small}
\begin{center}
\begin{tabular}{lccccl}
\hline \hline

Region  & Number of Detections & Sy1s (\%) & HBLR Sy2s (\%) &
non-HBLR
Sy2s (\%) & Starbursts (\%) \\

(1)&(2)&(3)&(4)& (5)& (6)\\
 \hline
I          &  23   & 61    & 30   &  9      &  0    \\
II         &  14   &36     &29    &  21     &14     \\
III        &  33   &  18   & 12   &  49     & 21    \\
IV         &  19   &  5    &5     &  21     & 69    \\
I+II       &  37   &  51   &  30  &  14     &  5    \\
III+IV     &  52   &  14   & 10   &  38     & 38    \\
I+III      &  56   &  36   &  20  &  32     &  12   \\
II+IV      &  33   &  18   &   15 &  21     & 46    \\
\hline

\end{tabular}
\parbox{6.5in}
{\baselineskip 9pt \noindent \vglue 0.5cm {\sc Note}: Column 1:
parameters. Column 2: number of sources. Column 3: the percents in
Sy1s. Column 4: the percents in HBLR Sy2s. Column 5: the percents in
non-HBLR Sy2s. Column 6: the percents in starburst galaxies.}
\end{center}
\end{small}
\end{table*}


Although the starburst galaxies perhaps show a cleaner separation
(less mixing), the regions of Figure 2 where Sy1s/HBLR Sy2s/non-HBLR
Sy2s mix are slightly significant. According to the unified model of
AGNs (Antonucci 1993) and some results (e.g., Tran 2003), it is
highly probable that HBLR Sy2s are the counterparts of Sy1s. We can
achieve a cleaner separation between Sy1s/HBLR Sy2s and non-HBLR
Sy2s. The three regions in the panels of Figure 2 are delineated by
the short solid lines roughly perpendicular to the dotted
AGN/starburst lines. Using the similar method of Dale et al. (2006),
the boundaries are

~~~~~~~~~~~~~~~~~~~~~~~~~~~~~~~~~~~~~~~~~~~~~~~~~~~~~~~~~~~~~~~~~~~~~~~~~~~~~~~

$\rm log (\frac{[Ne~V] 14.32~\mu m}{[Ne~II]12.81~\mu m})= 10 log
[EW(6.2~\mu m~PAH)]+8.0,$

~~~~~~~~~~~~~~~~~~~~~~~~~~~~~~~~~~~~~~~~~~~~~~~~~~~~~~~~~~~~~~~~~~~~~~~~~~~~~~~

$\rm log (\frac{[Ne~V]14.32~\mu m}{[Ne~II]12.81~\mu m})= 6.0 log
[EW(6.2~\mu m~PAH)]+0.3,$

~~~~~~~~~~~~~~~~~~~~~~~~~~~~~~~~~~~~~~~~~~~~~~~~~~~~~~~~~~~~~~~~~~~~~~~~~~~~~~~

$\rm log (\frac{[O~IV] 25.89~\mu m}{[Si~II] 34.82~\mu m})= 10 log
[EW(6.2~\mu m~PAH)]+7.8,$

~~~~~~~~~~~~~~~~~~~~~~~~~~~~~~~~~~~~~~~~~~~~~~~~~~~~~~~~~~~~~~~~~~~~~~~~~~~~~~~

$\rm log (\frac{[O~IV] 25.89~\mu m}{[Si~II] 34.82~\mu m})= 5.0 log
[EW(6.2~\mu m~PAH)]+0.8,$

~~~~~~~~~~~~~~~~~~~~~~~~~~~~~~~~~~~~~~~~~~~~~~~~~~~~~~~~~~~~~~~~~~~~~~~~~~~~~~~

$\rm log (\frac{[O~IV] 25.89~\mu m}{[S~III]33.48~\mu m})= 10 log
[EW(6.2~\mu m~PAH)]+8.5,$

~~~~~~~~~~~~~~~~~~~~~~~~~~~~~~~~~~~~~~~~~~~~~~~~~~~~~~~~~~~~~~~~~~~~~~~~~~~~~~~

$\rm log (\frac{[O~IV] 25.89~\mu m}{[S~III]33.48~\mu m})= 6.0 log
[EW(6.2~\mu m~PAH)]+1.5,$

\begin{equation}
\end{equation}
for regions I-II, II-III, IV-V, V-VI, VII-VIII, and VIII-IX,
respectively. The short solid lines separate appropriately Sy1s/HBLR
Sy2s, non-HBLR Sy2s, and starburst galaxies. In Table 5, we provide
the population statistics for these regions. Regions I+IV+VII and
III+VI+IX are representative (at the $>81\%$ level) of Sy1s/HBLR
Sy2s and starburst galaxies, respectively, while regions II+V+VIII
comprise a some mixture of pure types and are representative (at the
$>$50\% level ) of non-HBLR Sy2s. Non-HBLR Sy2s exhibit relatively
small line ratios and large PAH EWs in Figure 2 (see Table 7), which
is consistent with that non-HBLR Sy2s have a quite fraction of
starburst components (Wu et al. 2011).

\begin{figure}
\centerline{\includegraphics[scale=0.45,angle=0]{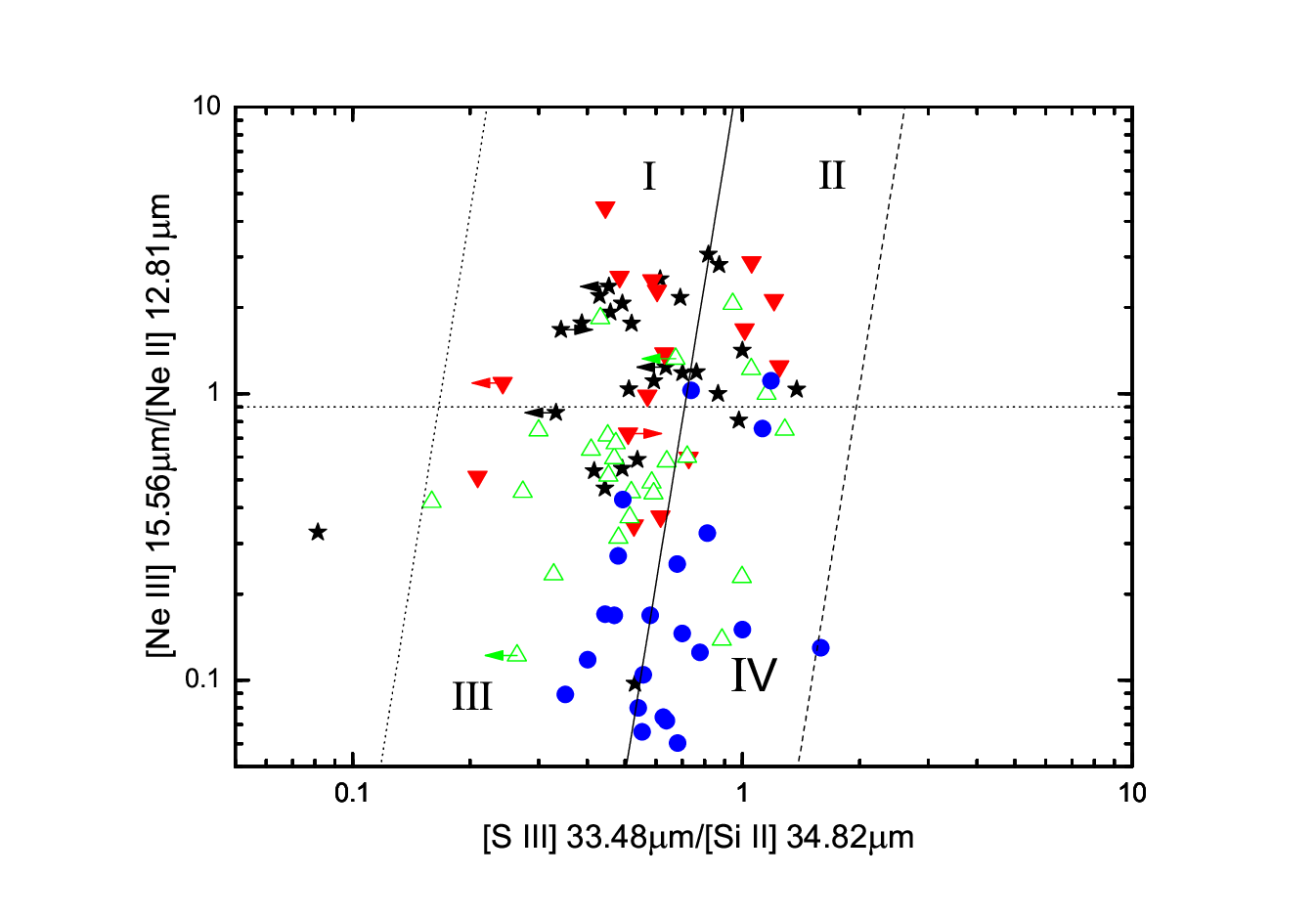}}
\caption{ Neon, Sulfur, and silicon diagnostic diagram involving
ratios of lines at different ionizations. The meaning of the symbols
is the same as Figure 2. The three lines are defined by the equation
7 and Roman numerals delineate regions distinguished by Sy1s, HBLR
Sy2s, non-HBLR Sy2s, and starburst galaxies (see table 6). [See the
electronic edition of the Journal for a color version of this
figure.]}
\end{figure}

The dotted lines in Figure 2 display a variational mix both an AGN
nucleus and a starburst region. We show the anchors for the mixing
model as follow:

~~~~~~~~~~~~~~~~~~~~~~~~~~~~~~~~~~~~~~~~~~~~~~~~~~~~~~~~~~~~~~~~~~~~~~~~~~~~~~~

$\rm EW ( \rm 6.2~\mu m~PAH) \approx 0.005~\mu m$

~~~~~~~~~~~~~~~~~~~~~~~~~~~~~~~~~~~~~~~~~~~~~~~~~~~~~~~~~~~~~~~~~~~~~~~~~~~~~~~

$\rm(\frac{[Ne v] 14.32~\mu m}{[Ne II]12.81~\mu m})\approx 1.0
(100{\%} ~AGN)$

~~~~~~~~~~~~~~~~~~~~~~~~~~~~~~~~~~~~~~~~~~~~~~~~~~~~~~~~~~~~~~~~~~~~~~~~~~~~~~~

$\rm EW( 6.2~\mu m~PAH) \approx 0.5~\mu m$

~~~~~~~~~~~~~~~~~~~~~~~~~~~~~~~~~~~~~~~~~~~~~~~~~~~~~~~~~~~~~~~~~~~~~~~~~~~~~~~

$\rm(\frac{[Ne v] 14.32~\mu m}{[Ne II]12.81~\mu m})\approx0.01
(0{\%} ~AGN)$

~~~~~~~~~~~~~~~~~~~~~~~~~~~~~~~~~~~~~~~~~~~~~~~~~~~~~~~~~~~~~~~~~~~~~~~~~~~~~~~

$\rm EW (\rm 6.2~\mu m~PAH) \approx 0.005~\mu m$

~~~~~~~~~~~~~~~~~~~~~~~~~~~~~~~~~~~~~~~~~~~~~~~~~~~~~~~~~~~~~~~~~~~~~~~~~~~~~~~

$\rm (\frac{[O IV] 25.89~\mu m}{[Si II] 34.82~\mu m})\approx 2.0
(100{\%} ~AGN)$

~~~~~~~~~~~~~~~~~~~~~~~~~~~~~~~~~~~~~~~~~~~~~~~~~~~~~~~~~~~~~~~~~~~~~~~~~~~~~~~

$\rm EW (6.2~\mu m~PAH) \approx 0.5~\mu m$

~~~~~~~~~~~~~~~~~~~~~~~~~~~~~~~~~~~~~~~~~~~~~~~~~~~~~~~~~~~~~~~~~~~~~~~~~~~~~~~

$\rm (\frac{[O IV] 25.89~\mu m}{[Si II] 34.82~\mu m})\approx0.02
(0{\%}~ AGN)$

~~~~~~~~~~~~~~~~~~~~~~~~~~~~~~~~~~~~~~~~~~~~~~~~~~~~~~~~~~~~~~~~~~~~~~~~~~~~~~~

$\rm EW (6.2~\mu m~PAH) \approx 0.005~\mu m$

~~~~~~~~~~~~~~~~~~~~~~~~~~~~~~~~~~~~~~~~~~~~~~~~~~~~~~~~~~~~~~~~~~~~~~~~~~~~~~~

$\rm(\frac{[O IV] 25.89~\mu m}{[S III] 33.48~\mu m})\approx 2.0
(100{\%} ~AGN)$

~~~~~~~~~~~~~~~~~~~~~~~~~~~~~~~~~~~~~~~~~~~~~~~~~~~~~~~~~~~~~~~~~~~~~~~~~~~~~~~

$\rm EW (6.2~\mu m~PAH) \approx 0.5~\mu m$

~~~~~~~~~~~~~~~~~~~~~~~~~~~~~~~~~~~~~~~~~~~~~~~~~~~~~~~~~~~~~~~~~~~~~~~~~~~~~~~

$\rm(\frac{[O IV] 25.89~\mu m}{[S III] 33.48~\mu m})\approx 0.01
(0{\%} ~AGN)$
\begin{equation}
\end{equation}

These results can be understood. The PAH emission is a tracer of
star forming regions in other galaxies (Laurent et al. 2000b;
Peeters et al. 2004; Brandl et al. 2006; Dale et al. 2006; Draine et
al. 2007; Smith et al. 2007), and its luminosity is in proportion to
the star formation rate in star forming galaxies (Roussel et al.
2001; Brandl et al. 2006; Calzetti et al. 2007; Shi et al. 2007;
Farrah et al. 2007; Baum et al. 2010). In Figure 2, we know that the
6.2$\mu$m PAH EW is a very useful discriminant between starburst
galaxies and Seyfert galaxies.

\begin{figure}
\centerline{\includegraphics[scale=0.45,angle=0]{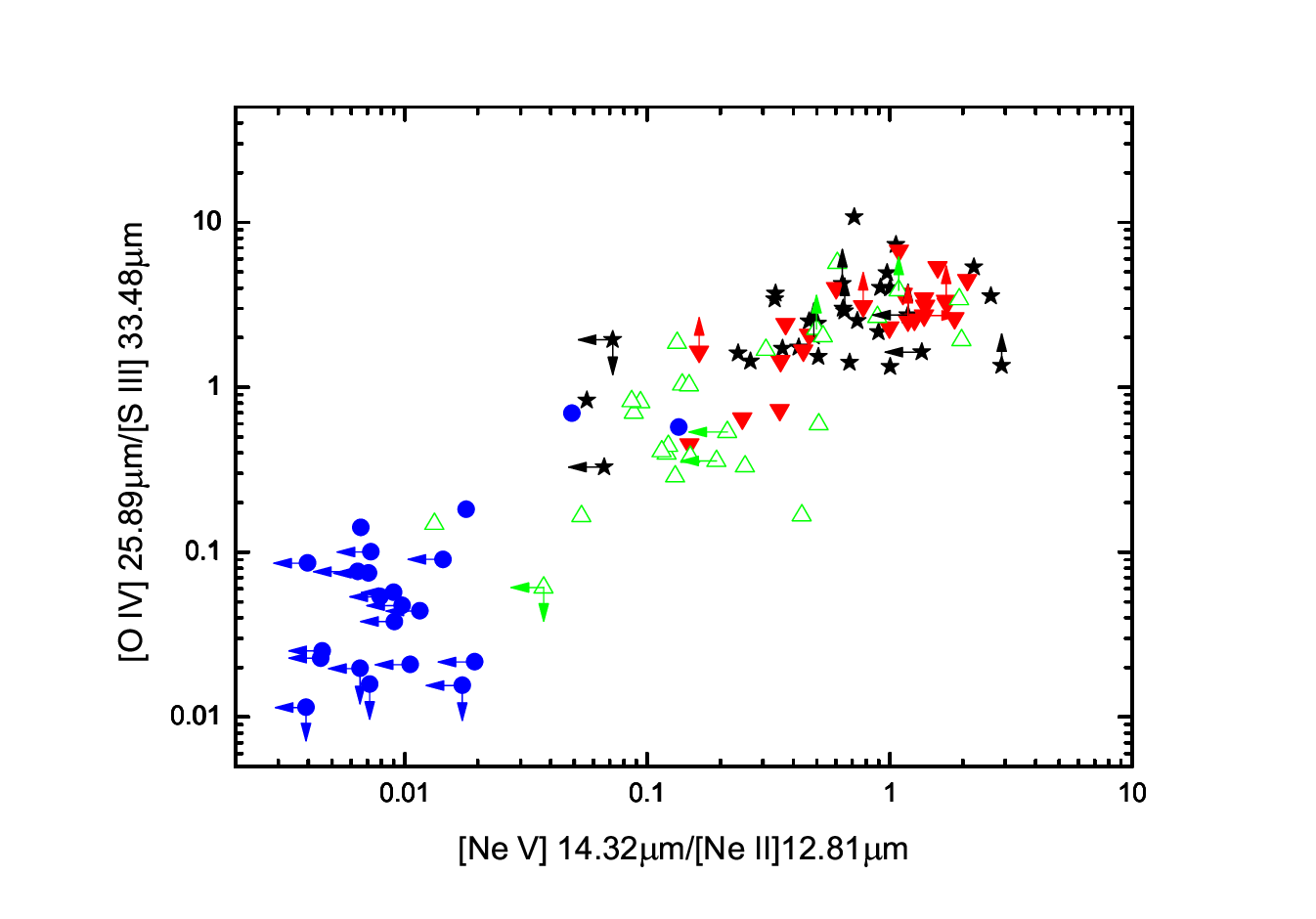}}
\caption{ \Nev $14.32~\mu m$/\NeII~$12.81~\mu m$ vs \oiv $25.89~\mu
m$/\NeII~ $12.81~\mu m$ for Seyfert and starburst galaxies in our
sample. The meaning of the symbols is the same as Figure 2. A clear
positive correlation is observed, representing an excitation
sequence anchored with the starburst galaxies, moving up in
excitation level to the non-HBLR and HBLR Sy2s, and ending with the
Sy1s. [See the electronic edition of the Journal for a color version
of this figure.]}
\end{figure}

\subsubsection{A Neon, Sulfur, and Silicon Diagnostic}\label{S:types}

In Figure 3, we use the similar method of Dale et al. (2006): the
neon excitation is regarded as a function of \Siii~$33.48~\mu m$/\Si
ii~$34.82~\mu m$, and the separation between the starburst and
AGN-powered sources is clear. \Neiii~$15.56~\mu m$/\Neii~$12.81~\mu
m$ line ratio is often utilized to interpret the excitation
properties of star formation galaxies (Thornley et al. 2000; Verma
et al. 2003; Brandl et al. 2006). Since producing $\rm Ne^{+}$ and
$\rm Ne^{++}$ needs the ionization potentials of 21.6 and 41.07 eV,
respectively, \Neiii/\Neii~ratio is sensitive to the most massive
stars in a starburst galaxy or to the presence of an AGN (Veilleux
et al. 2009). In Figure 3, starburst galaxies and non-HBLR Sy2s
exhibit a lower neon excitation while Sy1s and HBLR Sy2s exhibit a
higher neon excitation. Table 6 shows the source type fractions
within each of the four regions delineated by the lines drawn in
Figure 3. The boundaries are defined by lines with the same slope
but differing offsets (Dale et al. 2006):

\begin{figure}
\centerline{\includegraphics[scale=0.45,angle=0]{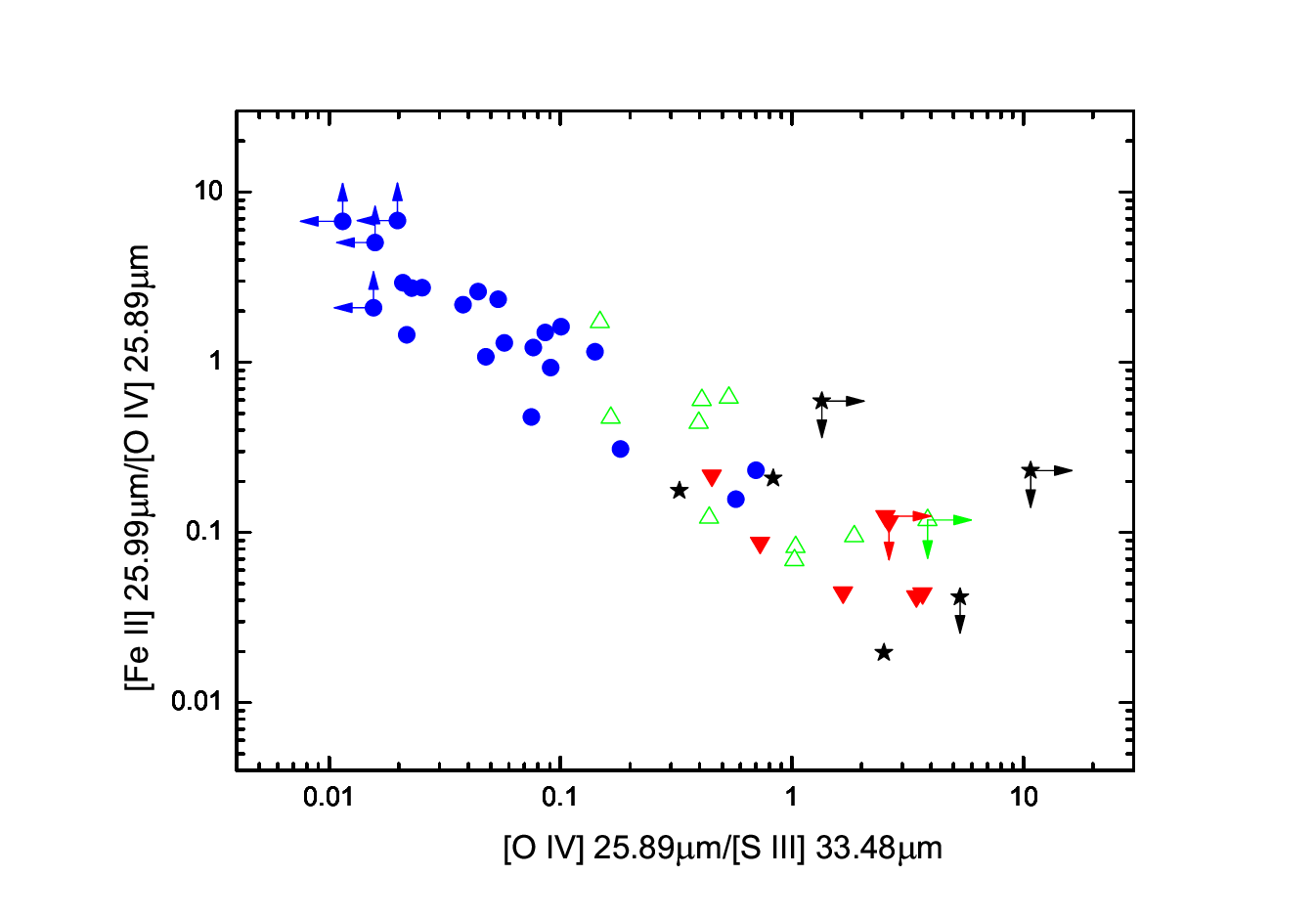}}
\caption{Mid-infrared diagnostic from Sturm et al. (2006). The
meaning of the symbols is the same as Figure 2. [See the electronic
edition of the Journal for a color version of this figure.]}
\end{figure}

\begin{equation}
\rm log (\frac{[Ne III] 15.56~\mu m}{[Ne II]12.81~\mu
m})=8.4log(\frac{[S III] 33.48~\mu m}{[Si II]34.82~\mu m})+\gamma,
\end{equation}
$\gamma=(+6.5, +1.2, -2.5)$ for the lines demarcating regions (I-II,
III-IV).

To estimate the statistical reliability of a classification for a
galaxy with the mid-infrared lines, we provide the numbers in Table
6 with the similar means of Dale et al. (2006).
In terms of the sample of Dale et al. (2006), this diagram can
provide a more significant separation. In Figure 3, Sy1s and HBLR
Sy2s mainly appear regions I and II while non-HBLR Sy2s and
starburst galaxies mostly reside in regions III and IV. We note,
however, that there is scatter in distributions of Sy1s, HBLR Sy2s,
and non-HBLR Sy2s.

In Figure 3, Sy1s and HBLR Sy2s will exhibit relatively strong \Si
ii~$34.82~\mu m$ emission and somewhat higher \Neiii~$15.56~\mu
m$/\Neii~$12.81~\mu m$ ratios. Non-HBLR Sy2s show relatively strong
\Si ii~$34.82~\mu m$ emission, while starburst galaxies display
strong signatures of \Siii~$33.48~\mu m$ and somewhat lower \Neiii~
$15.56~\mu m$/\Neii~$12.81~\mu m$ ratios. In Table 6, it is clear
that Region I and IV are representatives (at the $91\%$ and $69\%$
levels, respectively) of Sy1s/HBLR Sy2s and starburst galaxies.

According to the cooling line physics of Dale et al. (2006), these
results may be partially understood. The \Si ii~34.82 $\mu m$ line
is a significant coolant of X-ray-irradiated gas (Maloney et al.
1996). AGNs usually have stronger \Si ii~emission than starburst
galaxies, and when \oiv~is not easily detected, \Si ii presents an
obvious advantage (Hao et al. 2009). In addition, the \Siii
$33.48~\mu m$ line is regarded as a good marker of \HII~ regions
(Dale et al. 2006). In our subsamples, they are not distinguished
with the \Siii 33.48 $\mu m$/\Si ii 34.82 $\mu m$. The differences
in \Siii 33.48 $\mu m$/\Si ii 34.82 $\mu m$ among different
subsamples almost are not statistically significant (see Table 7).

In addition, we use the other fine structure line ratios to study
the separation of galaxies in our sample. The \oiv~25.89 $\mu
m$/\Siii~33.48 $\mu m$ and \Nev~14.32 $\mu m$/\Neii~12.81 $\mu m$
ratios are considered by us. Since the
 ionization potentials needed to product $\rm Ne^{4+}$ is 97.1 eV,
 this \Nev~line is detected in 28/44 Sy1s, 25/51 HBLR Sy2s,
  32/73 non-HBLR Sy2s, and 5/22 starburst galaxies.
 \oiv~25.89 $\mu m$ is also a good indicator of AGN activity (\oiv~
 25.89 $\mu m$ is not as good an AGN indicator as the PAH
 EW, e.g., contaminating \oiv~emission from WR
 stars and ionizing shocks, Lutz et al. 1998; Abel
 \& Saryapal 2008; Veilleux et al. 2009). This line is detected in
 33/44 Sy1s, 23/51 HBLR Sy2s, 34/73 non-HBLR Sy2s,
  and 18/22 starburst galaxies (the very low level \oiv~
  emission seen in starburst galaxies is not
generally attributed to AGN but to either supernova or wind-related
ionizing shocks or very hot stars; Lutz et al. 1998; Schaerer \&
Stasinska 1999). Figure 4 shows a more obvious
  separation of starburst and AGN-powered sources. The differences
   in \oiv~25.89 $\mu m$/\Siii~33.48 $\mu m$ and
   \Nev~14.32 $\mu m$/\Neii~12.81 $\mu m$ ratios among different
   subsamples are statistically significant (see Table 7).

\begin{figure}
\centerline{\includegraphics[scale=0.45,angle=0]{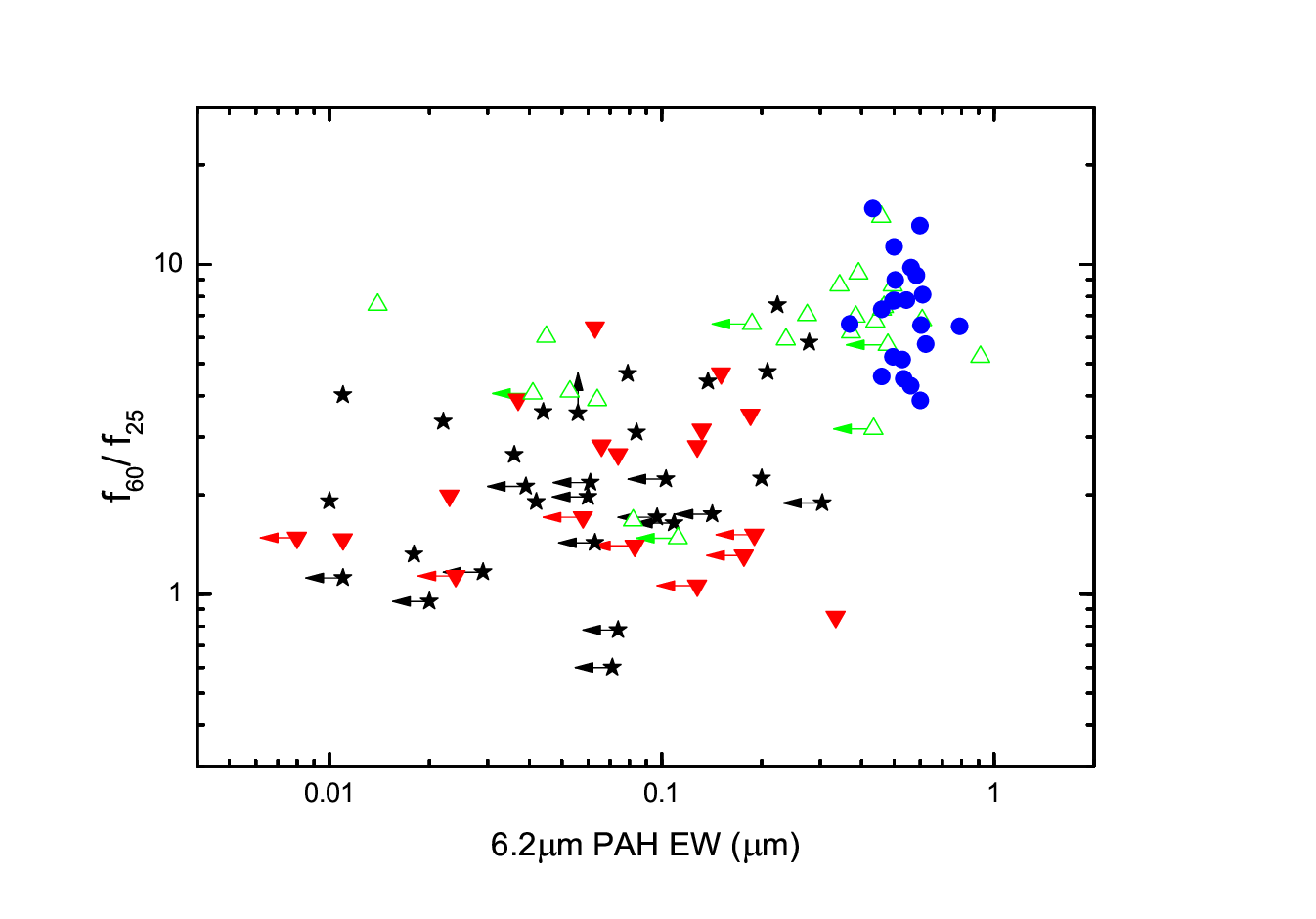}}
\caption{The $IRAS~f_{60}/f_{25}$ flux ratio as a function of the
$6.2~\mu m$ PAH feature equivalent width. The meaning of the symbols
is the same as Figure 2. [See the electronic edition of the Journal
for a color version of this figure.]}
\end{figure}

\begin{figure*}
\begin{center}
\includegraphics[width=18cm,height=11cm]{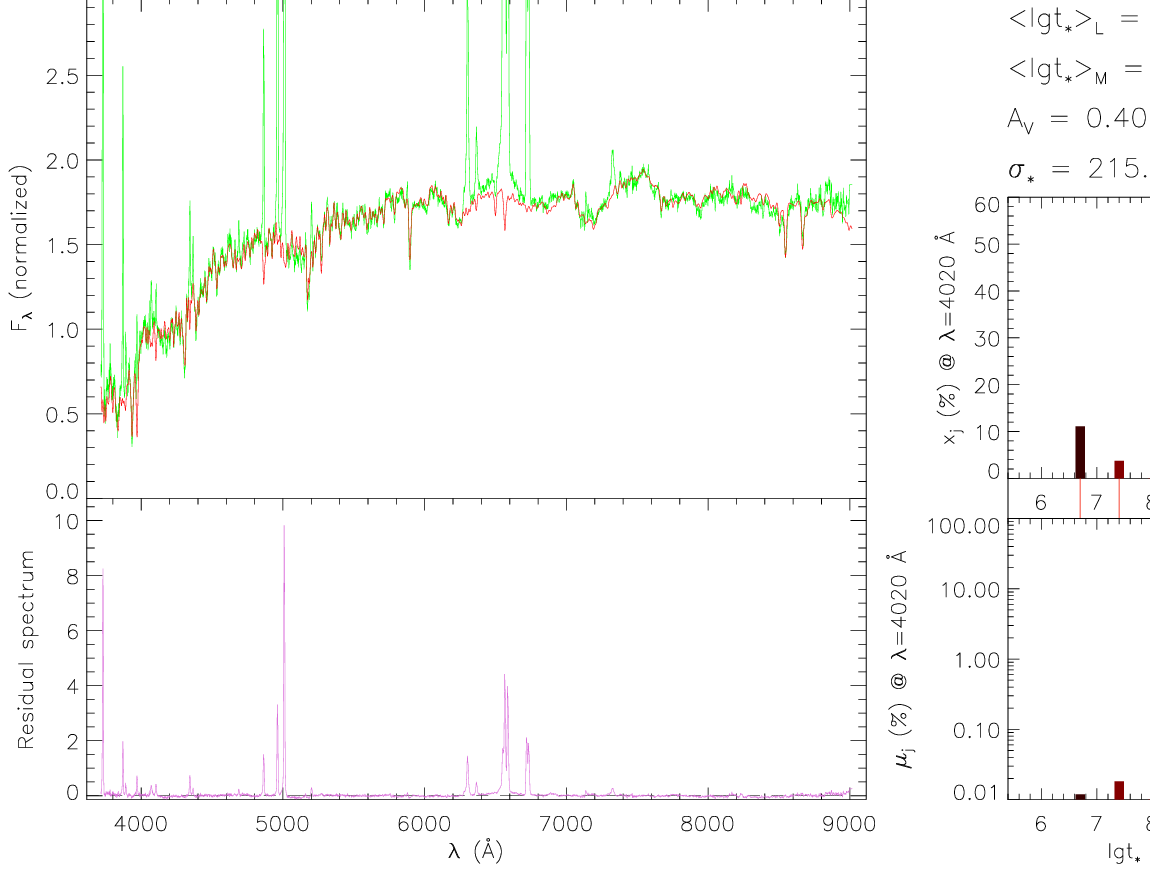}
\caption{Spectral synthesis of representative in Sy1 or HBLR Sy2
samples. Left top panel: the observed spectrum $O_{\lambda}$ (green)
and the model spectrum $M_{\lambda}$ (red). Left bottom panel: the
residual spectrum $E_{\lambda}$ (purple). Right: light (top) and
mass (bottom) weighted stellar population fractions $x_{\rm j}$ and
$u_{\rm j}$, respectively. The inserted panel on the right marks the
ages of the stellar population templates. The flux intensities of
the left two panels are normalized at $4020 \mathring{\rm A}$.}
\end{center}
\end{figure*}

\begin{figure*}
\begin{center}
\includegraphics[width=18cm,height=11cm]{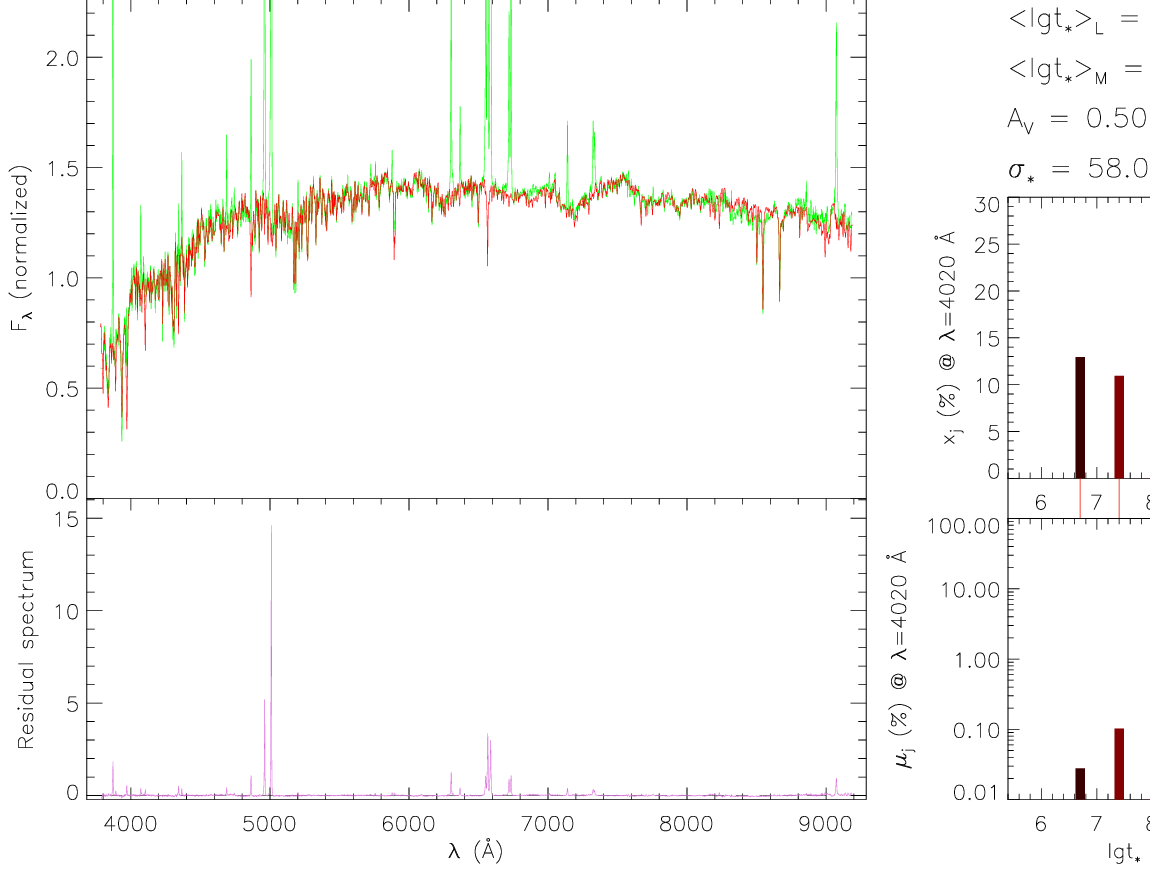}
\caption{Spectral synthesis of representative in non-HBLR Sy2 samples.
The related description in this Figure is the same as Figure 7.}
\end{center}
\end{figure*}

In Figure 5, we investigate the positions of galaxies on the \Feii~
25.99 $\mu m$/\oiv~25.89 $\mu m$ versus \oiv~25.89 $\mu m$/\Siii~
33.48 $\mu m$ diagnostic diagram. In the diagram, starburst galaxies
are preferentially in upper left corner while HBLR Sy2s/Sy1 have the
trend of residing in lower right corner. This diagram can separate
well starburst galaxies from those dominated by AGNs. There is
scatter in Sy1s distribution, we note, however, the trend may not be
change.

These results may be partially understood in the context of \Feii~
introduced below. The \Feii~25.99 $\mu m~$ and \oiv~25.89 $\mu m$
lines are observed in starbursts and supernova remnants. In
partially ionized zones of shocks, \Feii~is adequately emitted (e.g.,
Graham et al. 1987; Hollenbach \& McKee 1989) and again boosted by
shock destruction of grains (e.g., Jones et al. 1996; Oliva et al.
1999a, b) onto which Fe is normally depleted (Lutz et al. 2003). In
AGN, compared to the strong \oiv~emission from the NLR, it is much
stronger than the \Feii~emission from shocked or UV/X-ray irradiated
partially ionized zones (Lutz et al. 2003).

In Figure 6, we show the $IRAS~f_{60}/f_{25}$ flux ratio versus 6.2
$\mu m$ PAH EW. The difference in color between starburt and Seyfert
galaxies may be due to the relative strength of the host galaxies
and nuclear emissions (Alexander 2001), so Wu et al. (2011)
suggested that the $f_{60}/f_{25}$ ratio denotes the relative
strength of starbursts and AGN emissions. We find a better
correlation of the $IRAS~f_{60}/f_{25}$ ratio versus 6.2 $\mu m$ PAH
EW with Pearson's correlation coefficients of 0.59 and a probability
of $<$ 0.0001. HBLR Sy2s and Sy1s preferentially appear lower left
corner of this diagram while starburst galaxies and non-HBLR Sy2s
reside in upper right corner of this diagram. Our results are
consistent with those expected for that the $IRAS~f_{60}/f_{25}$
flux ratio and EW(6.2 $\mu m$ PAH) are an AGN/SB discriminator.

These ratios can distinguish between starburst galaxies, non-HBLR
Sy2s, and HBLR Sy2s/Sy1s. Based on the presentation of Section 2, we
can exclude that they could arise from a geometric effects
(orientation and solid-angle obscuration). So we suggest that these
ratios should show a difference in dominant mechanisms among these
galaxies.

\subsection{The Probability of Their Evolutionary Sequence}\label{S:types}

\begin{figure*}
\begin{center}
\includegraphics[width=18cm,height=11cm]{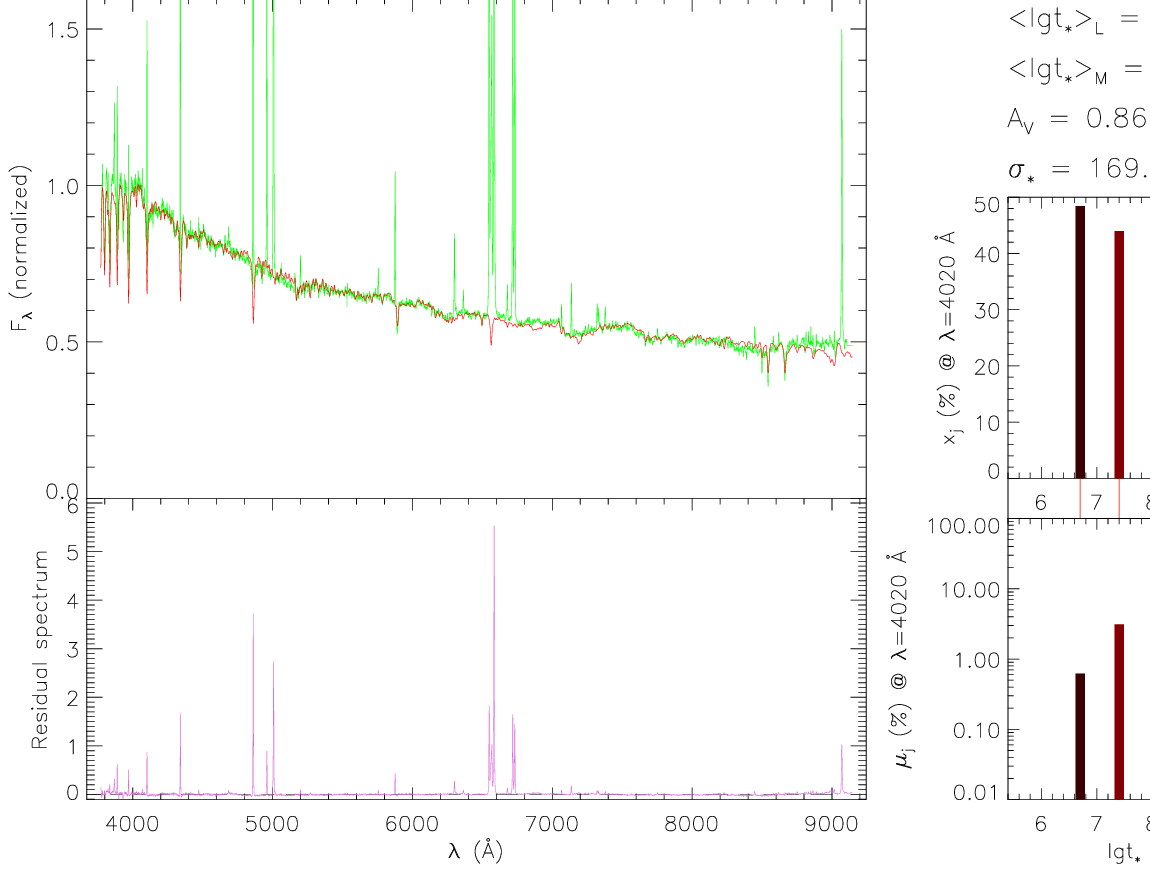}
\caption{Spectral synthesis of representative in starburst samples. The
related description in this Figure is the same as Figure 7.}
\end{center}
\end{figure*}

\subsubsection{Spectral Synthesis}\label{S:types}

To unveil both the star formation history and
the evolution of galaxies, grasping the
overall stellar populations of galaxies is vital and essential.
Based on the characteristics of circumnuclear stellar populations,
Storchi-Bergmann et al. (2001) proposed the evolutionary scenario
for Sy2s. To investigate the probability of evolution between
starburst and Seyfert galaxies, we utilize their stellar populations
to trace their possible evolution.

Since HBLR Sy2s could be the counterparts of Sy1s (Tran 2003), we
put HBLR Sy2s into Sy1 subsample. We search all objects of our
sample observed by the SDSS and find 6/4, 10, and 3 observed objects
which represent Sy1/HBLR Sy2, non-HBLR Sy2, and starburst
subsamples, respectively. To derive stellar populations of each of
our subsamples, we utilize the spectral synthesis code  \starlight\
(Cid Fernandes et al. 2005), which fits the observed spectrum with
template spectra of stellar populations and gives the light and mass
fractions of each stellar population. \starlight\ also gives the velocity dispersion
($\sigma_{\star}$) of the central region, reddening term $\rm A_{\nu}$, etc.
The accuracy and uncertainty of the fitting result had been tested and
given by Cid~Fernandes et al. (2004, 2005).
They combined stellar populations by their ages into ``young" ($t_j < 10^8$\,yr), ``intermediate-age"
($10^8 \leq t_j \leq 10^9$\,yr) and ``old" ($t_j > 10^9$\,yr, $t_j$ is the age of
stellar population) populations, and found that as the
signal-to-noise ratio (S/N) increase from 10 to 20, the uncertainty of the young population
decreases from 5\% to 3\%, and the other two populations decrease from 10\% to 6\%,
which suggest that the condensed populations can be well recovered by \starlight\ .

The observed spectra were taken from SDSS Data Release 7.
The spectral wavelength coverage is about 3800-8900~\AA.
All spectra of the subsamples have been corrected for Galactic reddening using dust maps from
Schlegel et al. (1998) and also have been corrected for redshift effect.
Since the fiber aperture of SDSS spectrographs are 3\arcsec, covering the nuclear region,
so the fluxes of spectra are mostly come from the bulges of the galaxies.

We adopt simple stellar
populations (SSPs) spectra from the Bruzual $\&$ Charlot (2003, hereafter
BC03) evolutionary synthesis models as our template spectra and
adopt 3 metallicities (Z=0.4, 1, and 2.5$\rm Z_{\sun}$)
and 10 ages (0.005, 0.025, 0.1, 0.29, 0.5, 0.9, 1.4, 2.5,
4, and 11 Gyr), which were computed with ``Padova 1994" evolutionary
tracks (Alongi et al. 1993; Bressan et al. 1993; Fagotto et al.
1994a,b; Girardi et al. 1996) and Chabrier (2003) initial mass
function. The metallicities adopted here are representative of neaby galaxies (Gallazzi et al. 2005).

Emission lines \OII ~$\lambda \lambda$3726+3729,
\NeI ~$\lambda$3869, \oiii ~$\lambda \lambda$4959+5007, \HeI ~$\lambda$5876, \OI ~$\lambda$6300,
 \NII ~$\lambda \lambda$6548+6583, \SII ~$\lambda \lambda$6716+6731,
H$\alpha$, H$\beta$, \NaI~D5890 galactic interstellar medium absorption
line, etc.\ are all masked and excluded from fitting.

The subsamples selected for stellar population synthesis are those have
stellar absorption lines and have S/N\,$>$\,10.
The source numbers of the subsamples
(Sy1/HBLR Sy2 subsample, non-HBLR Sy2 subsample, and starburst subsample)
are 2/1, 9, and 2, respectively.
Figures 7, 8, and 9 present the probable stellar population components of the
3 subsamples given by \starlight\ spectral synthesis result.
In each figure, the {\it left-top} panel shows the observed SDSS spectrum ({\it green}) and
the model synthetic spectrum ({\it red});
the {\it left-bottom} panel gives the residual spectrum ({\it purple});
the {\it right-top} panel shows the light-weighted fractions $x_{\rm j}$ of the template
stellar populations with logarithmic ages $\lg t_*$ (yr);
the {\it right-bottom} panel shows the mass-weighted fractions $\mu_{\rm j}$.

Following Cid Fernandes et al. (2005), we calculate
mean ages weighted by the flux and stellar mass as characteristic ages of the nuclear regions:

$ \langle\rm lg
t_{\star}\rangle_{L}$=$\sum\limits_{j=1}^{N_{\star}}x_{\rm j}\rm lg
t_{\rm j}$,

where $x_{j}$ is the light-weighted fraction of each population, and

$ \langle\rm lg
t_{\star}\rangle_{M}$=$\sum\limits_{j=1}^{N_{\star}}u_{\rm j}\rm lg
t_{\rm j}$

where $\mu_{j}$ is the fraction of stellar mass contributed by each
SSP. Since the luminosity of a stellar population is very sensitive to young massive stars,
so $\langle \lg t_{\star} \rangle_{L}$ can trace recent star formation and evolution history
better than $\langle \lg t_{\star} \rangle_{M}$.
The statistical mean ages $\overline {\langle \lg t_{\star} \rangle_{L}}$
of the HBLR Sy2s/Sy1s, non-HBLR Sy2s, and starburst galaxies
subsamples are $10^{9.3\pm0.15}$, $10^{8.94\pm0.18}$, $10^{7.5\pm0.2}$ yr, respectively.
We suggest that, to some extent, these characteristic ages show an evolution sequence of
the subsamples.

\subsubsection{The Possible Evolution}\label{S:types}

In Section 4.3.1, we show the age sequence of stellar populations
from starburst galaxies to non-HBLR Sy2s and then to HBLR Sy2s/Sy1s.
In fact, there are some works to represent indications.
Koulouridis et al. (2006) suggested that when the interaction is
getting weaker, starburst activity is reduced and an obscured AGN
appears simultaneously. Moreover, Wu et al. (2011) found that the
non-HBLR Sy2s and HBLR Sy2s are dominated by starbursts and AGNs,
respectively, and that there is a list from larger obscuration
(non-HBLR Sy2s) to smaller obscuration (HBLR Sy2s). Based on the
above analysis, we investigate the probability of evolution sequence
from starburst galaxies to non-HBLR Sy2s and then to HBLR Sy2s/Sy1s
with another means.

From a purely observational point of view, the BPT diagram is the
one of the best distinguishable means for starburst and Seyfert
galaxies. We introduce the evolution of galaxies that lie below the
dashed Ka03 line. Star-forming galaxies form a tight sequence from
low metallicities (low \NII/H\a, high \oiii/H\b) to high
metallicities (high \NII/H\a, low \oiii/H\b), which were regarded as
the ``star-forming sequence" (Kewley et al. 2006). The high
metallicity end of the star-forming sequence is the beginning of the
 AGN mixing sequence, and the sequence is extended towards high
 \oiii/H\b~ and \NII/H\a~ values
(Kewley et al. 2006; Yuan et al. 2010). This shows that the
\NII/H\a~ ratio can display well the evolutionary relation in star-forming
 galaxies, and it may also demonstrate that in some galaxies.

The nitrogen is one of the most abundant elements in the universe, but its nucleosynthesis origin is a long-term puzzle. In term of a primary nucleosynthesis, nitrogen is synthesized by fresh carbon generated by the parent star during hydrogen (H)-burning; in terms of a secondary nucleosynthesis, nitrogen should be synthesized by both carbon and oxygen initially present in the parent star during H-burning, and its abundance is proportional to the initial heavy element abundance (Matteucci 1986).
It appears that the abundances of carbon, nitrogen, and
oxygen can provide a helpful test for the stellar nucleosynthesis and galactic evolution models.

Based on the chemical evolutionary point of view, the dwarf irregular and blue compact dwarf (BCD) galaxies, the simplest objects assumed commonly, were dealt with in accordance with their chemical evolution in some works (Lequeux et al. 1979; Matteucci \& Chiosi 1983; Matteucci \& Tosi 1985; Vigroux et al. 1987; Garnett 1990). Although the theory of these galaxies' chemical evolution is far from being established (Pilyugin 1993), we have achieved some progresses in the related study, and Chiappini et al. (2003) presented the chemical evolution models for dwarf irregulars and spirals, which well reproduce the available constraints for these objects under the supposition that the stellar nucleosynthesis should be the same for all galaxies.  Next we demonstrate the possible evolutions in our subsamples.

In Table 7, the mean values of \NII/H\a~ratios increase with the
sequence from starburst galaxies to non-HBLR Sy2s and then to HBLR
Sy2s. We know that nitrogen is produced during hydrogen burning via
the CN or CNO cycles and is known as a primary or secondary
element. Primary nitrogen production is largely independent of
metallicity and occurs predominantly in intermediate-mass (3-9
$M_{\sun}$) stars (Becker $\&$ Iben 1979, 1980; Renzini $\&$ Voli
1981; Matteucci \& Tosi 1985 ; Matteucci 1986); but, some primary
nitrogen production derived from massive stars is considered to be dependent on
stellar mass and metallicity (Chiappini et al. 2005, 2006; Mallery
et al. 2007; Levesque et al. 2010).

Stasi$\rm \acute{n}$ska et al. (2006) suggested that starburst
galaxies are dominated by massive stars, and if the stellar
populations of circumnuclear in non-HBLR Sy2s are dominated by
intermediate-mass stars, then the increase of \NII/H\a~ from
starburst galaxies to non-HBLR Sy2s can be well understood. Because
different mass stars contribute during nitrogen producing the
different types of nitrogen (Chiappini et al. 2003), the \NII~ flux
of starburst galaxies is dominated by the secondary nitrogen while
that of non-HBLR Sy2s is dominated by the primary nitrogen. With the
evolutionary process, the total abundance of nitrogen increases from
starbutst galaxies to non-HBLR Sy2s, and the metallicity too. So the
\NII/H\a~ ratio increases with the list from starburst galaxies to
non-HBLR Sy2s.

In addition, we note that the mean values of the ratios decrease
with the list from non-HBLR Sy2s to HBLR Sy2s. The reason, we
suggest, may be \NII~ suppressed by a collisional de-excitation
process. It is generally accepted that the NLR is stratified that
the high-density and high-ionization gas is located close to the
nucleus, and low-density and low-ionization gas is in the outer part
of the NLR (Tran et al. 2000; Nagao et al. 2003). At the inner NLR,
lines with low critical densities, such as \NII, \SII, which have
the lower ionization potential, are suppressed by a collisional
de-excitation process, while lines with high critical densities and
recombination lines are unaffected (Zhang et al. 2008).
Due to this suppression, it directly causes that the \NII/H\a~
ratios decrease with the list from non-HBLR Sy2s to HBLR Sy2s.

At the same time, the \oiii/H\b~ ratio is sensitive to the ionizing
radiation field, and the ratio seems to increase with the sequence
from starburst galaxies to non-HBLR Sy2s and then to HBLR Sy2s in
Table 7. Therefore, the BPT diagram seems to have a hint of the evolutionary
sequence from starburst galaxies to non-HBLR Sy2s and then to HBLR
Sy2s/Sy1s.

\section{Discussions}\label{S:sed}

\subsection{Classifications of Nuclei}\label{S:sed}

In this section, we firstly discuss the separation between starburst
galaxies, non-HBLR Sy2, and HBLR Sy2 galaxies through the optical
diagnostic. Based on the various factors, such as, the
metallicity, abundance, the separation can be understood well. Then
we discuss the classification between starburst galaxies, non-HBLR
Sy2, HBLR Sy2, and Sy1 galaxies through the infrared spectral
diagnostic. Table 7 shows the mean values of various quantities and
indicates a remarkable separation between Sy1s, HBLR Sy2s, non-HBLR
Sy2s, and starburst galaxies utilizing the various methods.

\subsubsection{Optical Classification of Nuclei}\label{S:sed}

\def\oiii{\ifmmode [O {\sc iii}] \else [O {\sc iii}]\ \fi}
\def\Neii{\ifmmode [Ne {\sc ii}] \else [Ne {\sc ii}]\ \fi}
\def\Neiii{\ifmmode [O {\sc iii}] \else [Ne {\sc iii}]\ \fi}
\def\Nev{\ifmmode [O {\sc v}] \else [Ne {\sc v}]\ \fi}
\def\oiv{\ifmmode [O {\sc iv}] \else [O {\sc iv}]\ \fi}
\def\Feii{\ifmmode [Fe {\sc ii}] \else [Fe {\sc ii}]\ \fi}
\def\Siii{\ifmmode [S {\sc iii}] \else [S {\sc iii}]\ \fi}
\def\Si ii{\ifmmode [Si {\sc ii}] \else [Si {\sc ii}]\ \fi}

\begin{table*}
\caption{\small Statistical Properties of Starbursts, non-HBLR Sy2s,
HBLR Sy2s, and Sy1s}
\begin{small}
\begin{center}
\setlength{\tabcolsep}{1.0pt}
\renewcommand{\arraystretch}{1.5}
\begin{tabular}{lcccccccccl} \hline \hline
name& parameters  & $ \rm \frac{[O~III ] \lambda5007}{H\beta}$ &
$\frac{[\rm N~II]\lambda6584}{H\alpha}$ & $f_{60}/f_{25}$  &
\Nev/\Neii &\Neiii/\Neii& \oiv/\Siii & \Feii/\oiv
&  \Siii/\Si ii & EW  \\
      (1) & (2) &(3) & (4) & (5) & (6)&(7)& (8) & (9)& (10) & (11) \\
\hline

S1s(S1)      &Mean    &   ...   & ... &2.83$\pm$0.26&0.76$\pm$0.13&1.35$\pm$0.14&3.27+0.47&0.11$\pm$0.04&0.58$\pm$0.05&0.06$\pm$0.01  \\
             &Number  &   ...   & ...     &     45  &   30    &    32   &   33    &    6   &   27    &   29      \\
H S2s(S2)    &Mean    &9.54$\pm$0.94&0.83$\pm$0.05&2.52$\pm$0.19&0.99$\pm$0.12&1.68$\pm$0.22&3.04+0.35&0.09$\pm$0.02&0.67$\pm$0.08&0.08$\pm$0.02  \\
             &Number  &   19    &   19    &     42  &   23    &    23   &   23    &    7    &   16    &   18      \\
NH S2s(S3)   &Mean &6.93$\pm$0.69&1.27$\pm$0.12&5.20$\pm$0.36&0.38$\pm$0.09&0.75$\pm$0.10&1.41+0.32&0.42$\pm$0.14&0.59$\pm$0.06&0.30$\pm$0.05 \\
             &Number  &   30    &   30    &     54  &   29    &    29   &   27    &   11    &   25    &   24      \\
SBs(S4)      &Mean    &1.37$\pm$0.34&0.70$\pm$0.17&7.52$\pm$0.59&0.013$\pm$0.006&0.27$\pm$0.06&0.11$\pm$0.04&2.34$\pm$0.45&0.70$\pm$0.06&0.54$\pm$0.02 \\
             &Number  &   15    &    15   &     22  &    22    &    22   &    22   &    22   &    22   &   21      \\

$p_{\rm null}^{~\rm a}(\%)$&S1-S2&...&... &  71.20  &  9.75    & 27.57  &   74.56 &  75.41  &  32.33  &  39.81    \\
            &S1-S3    &   ...   &   ...   &  0.00   &  0.21    &  0.12  &   0.01  &  13.89  &  77.79  &  0.01     \\
            &S1-S4    &   ...   &   ...   &  0.00   &  0.00    &  0.00  &  0.00   &  0.09   &  14.97  &  0.00     \\
            &S2-S3    &   4.34  &   2.14  &  0.00   &  0.00    &  0.03  &  0.02   &  5.68   &  24.06  &  0.19     \\
            &S2-S4    &   0.00  &   0.93  &  0.00   &  0.00    &  0.00  &  0.00   & 0.01    &  74.50  &  0.00     \\
            &S3-S4    &   0.00  &   0.07  &  0.11   &  0.00    &  0.01  &  0.00   &  0.04   &  11.20  &  0.00     \\

\hline

\end{tabular}
\end{center}
{ \noindent \vglue 0.1cm {\sc Notes}: Column 1: the types of sources
and the probabilities. Column 2: the parameters are Mean, Median,
and Number values for the four type objects, perspectively. Columns
$3-4$: the flux ratios. Column 5: the $IRAS~f_{60}/f_{25}~line~
ratios$. Column $6-10$: the line ratios in \Nev~$4.32~\mu
m$/\Neii~$12.81~\mu m$, \Neiii~$15.56~\mu m$/\Neii~$12.81~\mu m$,
\oiv~$25.89~\mu m$/\Siii~$33.48~\mu m$, \Feii~$25.99~\mu
m$/\oiv~$25.89~\mu m$, and \Siii~$33.48~\mu m$/\Si ii~$34.82~\mu m$.
Column 11: the $6.2~\mu m$ PAH equivalent width (in $\mu m$). S1s, H
S2s, NH S2s, SBs denote Sy1s, HBLR Sy2s, non-HBLR Sy2s, and
starburst galaxies, respectively.

$^{\rm a}$ The probability $p_{\rm null}$ (in percent) for the null
hypothesis that the two distributions are drawn at random from the
same parent population. When there are censored data, we use Peto
$\&$ Peto Generalized Wilcoxon Test in ASURV.}
\end{small}
\end{table*}



In Section 4.1, the \oiii/H\b~ and \NII/H\a~ ratios can distinguish
between non-HBLR Sy2s, HBLR Sy2s, and starburst galaxies. The
distributions of \NII~\l 6584/H\a~ratios among them have significant
differences. A Kolmogorov-Smirnov (K-S) test shows that all the
probabilities for two subsamples of three groups to be extracted
from the same parent population are less than $5\%$ (see Table 7).
The results can be understood, because the \NII~\l 6584/H\a~ratio
correlates with both the metallicity and ionization parameter
(Veilleux \& Osterbrock 1987; Kewley et al. 2001a; Levesque et al.
2010). At lower metallicities, the \NII~flux is dominated by the
abundance of primary nitrogen (Chiappini et al. 2005; Mallery et al.
2007); at higher metallicities, the production of secondary nitrogen
becomes prevalent (Alloin et al. 1979; Mallery et al. 2007); carbon
and oxygen which originally present in the star synthesize together
secondary nitrogen which is thus proportional to abundance (Levesque
et al. 2010).

The difference (with a confidence level of $4.34\%$ ) in the \oiii
\l 5007/H\b~ratios between HBLR and non-HBLR Sy2s is presented. We
also find the significant differences (with a confidence level of
about $10^{\rm -2}$; we adopt the some upper limits as the measured
values, since ASURV could not deal with a case that contained both
upper and lower limits) in the \oiii \l 5007/H\b~ratios between
starburst galaxies and HBLR or non-HBLR Sy2s. These results can be
partial understood, because the \oiii \l 5007/H\b~ratio is sensitive
to the hardness of the ionizing radiation field (Baldwin, Phillips,
$\&$ Terlevich 1981) and metallicity (Kewley et al. 2001a).

The basic idea underlying this diagram proposed by Stasi$\rm
\acute{n}$ska et al. (2006) is that the emission lines in \HII~
regions are powered by massive stars, so the intensities of
collisionally excited lines of AGNs are much more than those of
starburst galaxies. It indicates that galaxies with AGNs should be
prefer in the upper right-hand side of starburst galaxies in the
diagram. Obviously, it is consistent with the known results.

\subsubsection{Infrared Spectral Diagnostics of Nuclei}\label{S:sed}

Except for the distributions of the \NeV/\Neii~and
\oiv/\Siii~ ratios between Sy1s and HBLR
Sy2s, Table 7 shows that all the distributions of the two ratios
between two subsamples have significant differences.
Following Voit (1992) and Spinoglio \& Malkan (1992), tracers of the
AGN in our sample are the high ionization lines \NeV~ and \oiv. With
the exception of young Wolf-Rayet stars (Schaerer \& Stasi$\rm
\acute{n}$ska 1999), the majority of stars do not produce enough
energetic photons to excite \NeV~and \oiv~(Baum et al. 2010). In
addition, the line ratios \NeV/\Neii~and \oiv/\Siii~have proved
empirically to provide a good separation between AGNs and
starburst galaxies (Genzel et al. 1998; Sturm et al. 2002; 2006).

The differences in \Neiii/\Neii~ratios between two subsamples also
are significant, and the K-S test and ASURV tests show that  all the
probabilities for the two subsamples to be extracted from the same
parent population are less than $0.2\%$ with the exception of both
Sy1s and HBLR Sy2s (see Table 7). Because of the rather large
difference in the ionization potentials of Ne$^{++}$ (41 eV) and
Ne$^+$ (22 eV), the \Neiii/\Neii~ratio is a good tracer of the
hardness of the interstellar radiation field (Thornley et al. 2000;
Wu et al. 2006). Since the extinction effects on \Neiii~ $15.56~\mu
m$ and \Neii~$12.81~\mu m$ are similar, the ratio is only weakly
affected by extinction (Wu et al. 2006), so the \Neiii/\Neii~ratio
is a good diagnostic.

Except for the distributions of the \Feii/\oiv~ ratios between Sy1s and HBLR
Sy2s, most of the distributions of
the ratios between two subsamples are different~(the detailed confidence
levels for the differences, see Table 7). \Feii
$25.99~\mu m$ is likely to from fast ionising shocks (Verma et al.
2003), and O'Halloran et al. (2006) suggested that \Feii $25.99~\mu
m$ emission has been linked primarily to supernova shocks. As
mentioned above, the \oiv~line is correlate with other AGN tracers,
so the \Feii/\oiv~ratio is a distinguishing tool.

There are little differences in the distributions of \Siii/\Si ii~
ratios between two subsamples. Especially ASURV test shows that a
confidence level of $77.79\%$ between non-HBLR Sy2s and Sy1s.
Although the exhibition of silicate dust features is speculated generally
in the tori, other type 1 AGNs do not show the silicate emission
(Sturm et al. 2005). \Si ii $34.82~\mu m$ line is believed to partly
originate in photodissociation regions (PDRs; Sternberg \& Dalgarno
1995) and the majority of \Si ii~emission comes from \HII~ regions
(Roussei et al. 2007). In addition, many or all of
the galaxies have the \Siii~emission, and it likely originates in
star-forming gas (Dudik et al. 2007). In Table 7, the mean values of the
\Siii/\Si ii~ratios are very close to each other among our
subsamples. This indicates that \Siii~and \Si ii~\ emissions may mainly
come from \HII~ regions of either starburst or Seyfert galaxies.

As a result, our subsamples can be distinguished by the optical and
infrared diagnostics. However, the Sy1s and HBLR Sy2s subsamples is
an exception. It may indicate that both the Sy1s and HBLR Sy2s come
from the same sample, and HBLR Sy2s may be the counterparts of Sy1s
at edge-on orientation,

\subsection{The Possible Evolution}\label{S:sed}

In this section, we firstly discuss various evolutionary scenarios
between Seyfert galaxies and starburst galaxies. Then we discuss
that the two models of tidal features and the spectroscopic ages of
merger-induced star-forming regions were used as clocks to set the
relative ages of the different samples, but they have not been very
successful. We also discuss using \NII~ as the probability of an
evolutionary clock.

A supermassive black hole is made through successive mergers among
starburst remnants (e.g., Norman \& Scoville 1988; Taniguchi et al.
1999; Ebisuzaki et al. 2001; Mouri \& Taniguchi 2002b). Galaxy
interactions and mergers are fundamental to galaxy formation and
evolution. Since galaxy interaction supplies the gas at the high
efficiency, although the companion is not recognizable, Seyfert
galaxies with circumnuclear starbursts are likely to be interacting
(Mouri \& Taniguchi 2004). With regard to the merger, the most
widely supported merger scenario is based on the Toomre (1977)
sequence in which two galaxies lose their mutual orbital energy and
angular momentum and then coalesce into a single galaxy. The
above-mentioned two actions can supply a high rate of gas required
by starbursts and AGNs.

Many works studied and discussed the galaxy evolution, for
example, the evolutionary scenario of Heckman et al. (1989) and
Osterbrock (1993) can be summarized: starburst galaxies $\rightarrow$
starburst-dominant Seyfert $\rightarrow$ AGN-dominant Seyfert/Sy1s
(Mouri \& Taniguchi 2002a).
Since starbursts in Seyfert galaxies are older than those in
classical starburst galaxies, Mouri \& Taniguchi (2002a) suggested
the evolutionary path of starburst $\rightarrow$ starburst-dominant
Seyfert $\rightarrow$ host-dominant Seyfert $\rightarrow$ LINER for
a late-type galaxy and another evolutionary path of
(starburst$\rightarrow$) AGN-dominant Seyfert $\rightarrow$
host-dominant Seyfert $\rightarrow$ LINER for an early-type galaxy.

Mergers and tidal interactions between galaxies have been studied
extensively since the pioneering work of Toomre $\&$ Toomre (1972).
Some authors have tried to use tidal features or the spectroscopic
ages of merger-induced star-forming regions as clocks to set the
relative ages of the different samples. A study of tidal debris
associated with 126 nearby red galaxies was presented by van Dokkum
(2005) and he concluded that the majority of today's most luminous
field elliptical galaxies were assembled at low redshift through
mergers of gas-poor, bulge-dominated systems. These ``dry" mergers
are consistent with the high central densities of elliptical
galaxies, their old stellar populations, and the strong correlations
of their properties.

Based on $Spitzer$ MIPS observations, Bai et al. (2010 ) presented
the mid-IR study of galaxy groups in the nearby universe and
suggested that if galaxy$-$galaxy interactions are responsible, then
the extremely low starburst galaxy fraction ($<1\%$) implies a short
timescale ($\sim0.1$ Gyr) for any merger-induced starburst stage.
The models of utilizing tidal features and the spectroscopic ages of
merger-induced star-forming regions depend strongly on identifying
the red tidal features at higher redshift and supposing that the
star-forming gas is isothermal, respectively. To date we know that
they have not been very successful.

The stellar population in Seyfert 2 is an old stellar content
($>$1Gyr) and has high metallicities (up to three solar; Vaceli et al.
1997), while starburst galaxies have a younger population and show
solar metallicities or lower. We analyze the ages of stellar
populations in starburst galaxy, non-HBLR Sy2, and HBLR Sy2
subsamples with \starlight\, respectively, and show them in Figures 7,
8, and 9. Despite the existence of only 14 SDSS-observed objects and
poor fittings, we present a probable trend that the ages of stellar
population increase with the list from starburst galaxies to
non-HBLR Sy2s and then to HBLR Sy2s/Sy1s. Our result is consistent
with the known results and we show roughly that HBLR Sy2s have older
stellar population than non-HBLR Sy2s. Therefore, our supposition
that the stellar populations of circumnuclear in non-HBLR Sy2s are
dominated by intermediate-mass stars could be right.

Zhang et al. (2008) presented that Seyfert 1 and Seyfert 2 galaxies
have different distributions on the \NII/H\a~ versus \oiii/H\b~
diagram and interpreted that it is due to the obscuration of an
inner dense NLR by the torus. We argue that this explanation could
be unlikely, because it may not account for the difference in
\NII/H\a~ between HBLR and non-HBLR Sy2s.
\NII~suppressed by a collisional de-excitation process, we suggest,
may be the reason of the decrease of the \NII/H\a~ ratios. Since
non-HBLR Sy2s and HBLR Sy2s are dominated by starbursts and AGNs,
respectively (Wu et al. 2011), the \NII~ flux in non-HBLR Sy2s could
be affected less by this collisional de-excitation process or not at
all. In addition, our sequence model is agree well with scenarios of
Heckman et al. (1989), Osterbrock (1993) and Mouri \& Taniguchi
(2002a). Certainly, our proposal that uses the metallicity in the
form of \NII~ as a time discriminator should be further verified by
other observations.

In Section 5.1, we discuss the classifications between starburst
galaxies, non-HBLR Sy2, HBLR Sy2, and Sy1 galaxies through the
optical and infrared spectral diagnostic, and we show remarkable
separations among them with statistics. In Section 5.2, utilizing
the contribution of nitrogen from the different-mass stars and the
action of \NII~ suppressed by a collisional de-excitation process,
we may account for the changes of \NII~ flux in starburst galaxies,
non-HBLR S2ys, and HBLR S2ys.

\section{Summary}\label{S:sed}

We have carried out the BPT diagram and a detailed study of the
mid-infrared emission line properties, which derive from the
carefully selected sample of 45 Sy1s, 46 HBLR Sy2s, 57 non-HBLR
Sy2s, and 22 starburst galaxies. Using the optical diagnostic, the
mid-infrared diagnostic, and PAH EW, we can distinguish among our
subsamples. In addition, we also investigate the possible evolution
among them. The main results are the followings:

\begin{enumerate}

\item We find that the line, $\rm log (\frac{[O III]{5007}}{H\beta})=
  3.2\times log ([N II]/ H\alpha)+1.15$, can well distinguish between
  non-HBLR Sy2s and HBLR Sy2s on the BPT diagram.

\item The \oiii/H\b~ versus \NII/H\a~ diagram can separate Seyfert and
 starburst galaxies, which is consistent with the known results.

\item In our subsamples, starburst and Seyfert galaxies can be separated by
$6.2~\mu m$ PAH EW versus the \NeV~$14.32~\mu m$/\Neii~$12.81~\mu
m$, \oiv~$25.89~\mu m$/\Si ii~$34.82~\mu m$,
  and \oiv~$25.89~\mu m$/\Siii~$33.48~\mu m$ ratios, respectively.

\item The diagram, \Siii~$33.48~\mu m$/\Si ii~$34.82~\mu m$ against
  \Neiii~$15.56~\mu m$/\Neii~$12.81~\mu m$, can roughly distinguish
among our subsamples.

\item On the basis of statistics, the result that HBLR Sy2s may be the counterparts of
Sy1s at edge-on orientation also is consistent with the known
results.

\item The \oiii/H\b~ versus \NII/H\a~ diagram seems to intimate a evolutionary sequence from
 starburst galaxies to non-HBLR Sy2s and then to HBLR Sy2s/Sy1s.

\end{enumerate}

From these results we draw the following two main conclusions:

\begin{enumerate}

\item The BPT diagram and mid-infrared emission line ratios can distinguish or diagnose
  well the starburst galaxies, HBLR Sy2s, non-HBLR Sy2s, and Sy1s.

\item A evolutionary sequence of starburst galaxies, non-HBLR Sy2s,
  HBLR Sy2s, and Sy1s is presented and can be confirmed by using the changes of nitrogen fluxes.

\end{enumerate}

~~~~~~~~~~~~~~~~~~~~~~~~~~~~~~~~~~~~~~~~~~~~~~~~~~~~~~~~~~~~~~~~~

\acknowledgements
We thank the anonymous referee for the very good suggestions/comments
that significantly improved this paper. We also thank Y.-C. Liang for
helpful discussions. This research has made use of the
\spitzer~ data, the NASA/IPAC Extragalactic Database, and \MPA-JHU
Database.


\begin{thebibliography}

\bibitem[Abel \& Saryapal (2008)]{2008ApJ... 678.686}
Abel, N. P., $\&$ Saryapal, S. 2008, \apj, 678, 686

\bibitem[Adelman-McCarthy et al. (2006)]{2006ApJ...162.38}
Adelman-McCarthy, J. K., et al. 2006, \apjs, 162, 38

\bibitem[Alexander (2001)]{2001MNRAS...320.L15
}Alexander, D. M. 2001, \mnras, 320, L15

\bibitem[Alloin, Collin-Souffrin, \& Vigroux (1979)]{1979AA...78.200}
Alloin, D., Collin-Souffrin, S., Joly, M., \& Vigroux, L.
1979, \aap, 78, 200

\bibitem[Alongi et al. (1993)]{1993AA...97.851}
Alongi, M., Bertelli, G., Bressan, A., Chiosi,
C., Fagotto, F., Greggio, L., \& Nasi, E. 1993, \aaps, 97, 851

\bibitem[Antonucci (1993)]{1993ARAA...31.473}Antonucci, R. 1993, \araa, 31, 473

\bibitem[Armus et al. (1989)]{1989ApJ...347.727}Armus, L., et al. 1989, \apj, 347, 727

\bibitem[Armus et al. (2004)]{2004ApJ...154.178}Armus, L., et al. 2004, \apjs, 154, 178

\bibitem[Bai et al. (2010)]{2010ApJ...713.637}Bai, L., et al. 2010, \apj, 713, 637

\bibitem[Baldwin, Phillips, \& Terlevich (1981)]{1981PASP...93.5}
Baldwin, J. A., Phillips, M. M., \& Terlevich, R. 1981, \pasp, 93, 5

\bibitem[Baum et al. (2010)]{2010ApJ...710.289}Baum, S. A., et al. 2010, \apj, 710, 289

\bibitem[Becker \& Iben (1979)]{1979ApJ...232.831}Becker, S. A., \& Iben, I., Jr. 1979, \apj, 232, 831

\bibitem[Becker \& Iben (1980)]{1980ApJ...237.111}Becker, S. A., \& Iben, I., Jr. 1980, \apj, 237, 111

\bibitem[Bernard-Salas et al. (2009)]{2009ApJ...184.230}Bernard-Salas, J., et al. 2009, \apjs, 184, 230

\bibitem[Bernardi et al. (2003)]{2003ApJ...125.1817}Bernardi, M., et al. 2003, \aj, 125, 1817

\bibitem[Brandl et al. (2006)]{2006ApJ...653.1129}Brandl, B. R., et al. 2006, \apj, 653, 1129

\bibitem[Bressan et al. (1993)]{1993ApJ...100.647}Bressan, A., Fagotto, F., Bertelli, G., Chiosi, C.
1993, \aaps, 100, 647

\bibitem[Brightman \& Nandra (2008)]{2008MNRAS...390.1241}Brightman, M., \& Nandra. K. 2008, \mnras, 390, 1241

\bibitem[Brinchmann et al. (2004)]{2004MNRAS...351.1151} Brinchmann, J., Charlot, S., et al. 2004, \mnras, 351, 1151

\bibitem[Bruzual \& Charlot (2003)]{2003MNRAS344.1000}
Bruzual, G., $\&$ Charlot, S. 2003, \mnras, 344, 1000

\bibitem[Buchanan et al. (2006)]{2006AJ...132.401}
Buchanan, C. L., et al. 2006, \aj, 132, 401

\bibitem[Calzetti et al. (2007)]{2007ApJ...666.870}Calzetti, D., et al. 2007, \apj, 666, 870

\bibitem[Cid~Fernandes et al. (2004)]{2004MNRAS...355.273} Cid~Fernandes, R., Gu, Q., Melnick, J., Terlevich, E., Terlevich, R., Kunth, D., Rodrigues Lacerda, R., \& Joguet, B.\ 2004, \mnras, 355, 273

\bibitem[Cid~Fernandes et al. (2005)]{2005MNRAS...358.363} Cid~Fernandes R., Mateus A., Sodr\'e L., Stasi\'nska G., Gomes J. M., 2005, \mnras, 358, 363

\bibitem[Chabrier (2003)]{2003PASP...115.763}Chabrier, G. 2003, \pasp, 115, 763

\bibitem[Chiappini et al. (2006)]{2006AA...449.27}Chiappini, C., Hirschi, R., Meynet, G., Maeder,
A., \& Matteucci, F. 2006, \aap, 449, 27

\bibitem[Chiappini et al. (2005)]{2005AA...437.429}
Chiappini, C., Matteucci, F., \& Ballerro, S. k. 2005, \aap, 437,
429

\bibitem[Chiappini et al. (2003)]{2003MNRAS...339.63}Chiappini, C., Romano, D., \& Matteucci, F. 2003,
\mnras, 339, 63

\bibitem[Cid Fernandes et al. (2005)]{2005MNRAS...358.363}Cid Fernandes, R., Mateus, A., Sodr$\rm
\acute{e}$, L., Stasi$\rm \acute{n}$ska, G., \& Gomes, J. M. 2005,
\mnras, 358, 363

\bibitem[Constantin et al. (2009)]{2009ApJ...705.1336}Constantin, A., et al. 2009, \apj, 705, 1336

\bibitem[Contini et al. (1998)]{1998AAS130.285}Contini, T., Consider, S., \& Davoust, E. 1998, \aaps, 130, 285

\bibitem[Dale et al. (2006)]{2006ApJ...646.161}Dale, D. A., et al. 2006, \apj, 646, 161

\bibitem[Dale et al. (2009)]{2009ApJ...693.1821}Dale, D. A., Smith, J. D. T., Schlawin, E. A., et al. 2009, \apj, 693, 1821

\bibitem[Dasyra et al. (2008)]{2008ApJ...674.L9}Dasyra, K. M., et al. 2008, \apj, 674, L9

\bibitem[Deo et al. (2007)]{2007ApJ...671.124}Deo, R. P., et al. 2007, \apj, 671, 124

\bibitem[Deo et al. (2009)]{2009ApJ...705.14}Deo, R. P., Richards, G. T., Crenshaw, D. M., \&
Kraemer, S. B. 2009, \apj, 705, 14

\bibitem[Draine et al. (2007)]{2007ApJ...663.866}Draine, B. T., et al. 2007, \apj, 663, 866

\bibitem[Dudik et al. (2007)]{2007ApJ...664.71}Dudik, R. P.,
Weingartner, J. C., Satyapal, S., Fischer, J., Dudley, C. C., \&
O¡¯Halloran, B. 2007, \apj, 664, 71

\bibitem[Ebisuzaki et al. (2001)]{2001ApJ...562.L19}Ebisuzaki, T., et al. 2001, \apj, 562, L19

\bibitem[Fagotto et al. (1994a)]{1994aAA...104.365}Fagotto, F., Bressan, A., Bertelli, G., Chiosi, C.
1994a, \aaps, 104, 365

\bibitem[Fagotto et al. (1994b)]{1994bAA...105.29}Fagotto, F., Bressan, A., Bertelli, G., Chiosi, C.
1994b, \aaps, 105, 29

\bibitem[Farrah et al. (2007)]{ApJ...667.149}Farrah, D., et al. 2007, \apj, 667, 149

\bibitem[Gallazzi et al. (2002)] {2005MNRAS...362.41} Gallazzi, A., Charlot, S., Brinchmann, J., White, S.
D. M. 2005, \mnras, 362, 41

\bibitem[Gallimore et al. (2010)]{2010ApJS...187.172}
Gallimore, J. F., et al. 2010, \apjs, 187, 172

\bibitem[Garnett (1990)]{1990ApJ...363.142} Garnett, D. R. 1990, \apj, 363, 142

\bibitem[Genzel et al. (1998)]{1998ApJ...498.579}Genzel, R., et al. 1998, \apj, 498, 579

\bibitem[Girardi et al. (1996)]{1996AA...117.113}Girardi, L., Bressan, A., Chiosi, C.,
Bertelli, G., Nasi, E. 1996, \aaps, 117, 113

\bibitem[Goulding \& Alexander (2009)]{2009MNRAS...398.1165}Goulding, A. D., \& Alexander, D. M. 2009, \mnras, 398, 1165

\bibitem[Graham et al. (1987)]{1987ApJ...313.847}Graham, J. R., et al. 1987, \apj, 313, 847

\bibitem[Gu \& Huang (2002)]{2002ApJ...579.205}Gu, Q., \& Huang, J. 2002, \apj, 579, 205

\bibitem[Gu et al. (2001)]{2001ApJ...366.765}
Gu, Q., Maiolino, R., \& Dultzin-Hacyan, D. 2001, \aap, 366, 765

\bibitem[Gu et al. (2006)]{2006MNRAS...366.480}
Gu, Q., Melnick, J., Fernandes, R. C., et al. 2006, \mnras, 366, 480

\bibitem[Hao et al. (2009)]{2009ApJ...704.1159}Hao, L., Wu, Y. -L., et al. 2009, \apj, 704, 1159

\bibitem[Heckman et al. (1989)]{1989ApJ...342.735 }Heckman, T. M., Blitz, L.,
Wilson, A. S., Armus, L., \& Miley, G. K. 1989, \apj, 342, 735

\bibitem[Ho et al. (1997)]{1997ApJS...112.315}Ho, L. C., Filippenko, A. V., \& Sargent, W. L. W. 1997, \apjs, 112, 315

\bibitem[Ho (2008)]{2008ARAA46.475}Ho, L. C. 2008, \araa, 46, 475

\bibitem[Hollenbach \& McKee (1989)]{1989ApJ...342.306}Hollenbach, D., \& McKee, C. F. 1989, \apj, 342, 306

\bibitem[Hunt \& Malkan (1999)]{1999ApJ...516.660}Hunt, L. K., \& Malkan, M. A. 1999, \apj, 516, 660

\bibitem[Imanishi (2002)]{2002ApJ...569.44}Imanishi, M. 2002, \apj, 569, 44

\bibitem[Jones et al. (1996)]{1996ApJ...469.740}Jones, A. P., et al. 1996, \apj, 469, 740

\bibitem[Kauffmann et al. (2003a)]{2003aMNRAS...346.1055}Kauffmann, M. J., et al. 2003a, \mnras, 346, 1055

\bibitem[Kauffmann et al. (2003b)]{2003bMNRAS...341.54}Kauffmann, M. J., et al. 2003b, \mnras, 341, 54

\bibitem[Kaufman et al. (2006)]{2006ApJ...644.283}Kaufman, M. J., et al. 2006, \apj, 644, 283

\bibitem[Kewley et al. (2000)]{2000ApJ...530.704}Kewley, L. J., et al. 2000, \apj, 530, 704

\bibitem[Kewley et al. (2001a)]{2001aApJ...556.121}Kewley, L. J., Dopita, M., Sutherland, R., Heissler, C., \& Trevena, J. 2001a, \apj, 556, 121

\bibitem[Kewley et al. (2001b)]{2001bApJS...132.37}Kewley, L. J., et al. 2001b, \apjs, 132, 37

\bibitem[Kewley et al. (2006)]{2006MNRAS...372.961}Kewley, L. J., Groves, B., Kauffmann G., \& Heckman, T. 2006, \mnras, 372, 961

\bibitem[Koulouridis et al. (2006)]{2006ApJ...651.93}Koulouridis, E., et al. 2006, \apj, 651, 93

\bibitem[Koulouridis et al. (2002)]{2002ApJ...572.169}Krongold, Y., et al. 2002, \apj, 572, 169

\bibitem[Lagache et al. (2005)]{2005ARAA...43.727}Lagache, G., Puget, JL., \& Dole, H. 2005, \araa, 43, 727

\bibitem[Lamareille et al. (2004)]{2004,MNRAS...350.396}
Lamareille, F., Mouhcine, M., Contini, T., et al. 2004, \mnras,
350, 396

\bibitem[Laurent et al. (2000a)]{2000aAA...359.887}Laurent, O., et al. 2000a, \aap, 359, 887

\bibitem[Laurent et al. (2000b)]{2000bAA...654.L45}Laurent, O., et al. 2000b, \aap, 654, L45

\bibitem[Lequeux et al. (1979)]{1993AA...80.155}Lequeux, J., Peimbert, M., Rayo, J. F., Serrano, A., \& Torres-Peimbert, S. 1979, \aap, 80, 155

\bibitem[Levenson et al. (2001)]{2001ApJ...550.230}Levenson, N. A., et al. 2001, \apj, 550, 230

\bibitem[Levesque et al. (2010)]{2010AJ...139.712}Levesque, E. M., Kewley, L. J., \& Larson, K. L. 2010, \aj, 139, 712

\bibitem[Lumsden et al. (2004)]{2004MNRAS...348.1451}
Lumsden, S. L., Alexander, D. M., \& Hough, J. H. 2004,\mnras,
348, 1451

\bibitem[Lumsden et al. (2001)]{2001MNRAS...327.459}Lumsden, S. L., et al. 2001, \mnras, 327, 459

\bibitem[Lutz et al. (1998)]{1998AA...333.L75}Lutz, D., et al. 1998, \aap, 333, L75

\bibitem[Lutz et al. (2003)]{2003AA...409.867}Lutz, D., et al. 2003, \aap, 409, 867

\bibitem[Mallery et al. (2007)]{2007ApJS...173.482}Mallery, R. P., et al. 2007, \apjs, 173, 482

\bibitem[Maloney et al. (1996)]{1996ApJ...466.561}Maloney, P. R., et al. 1996, \apj, 466, 561

\bibitem[Matteucci \& Chiosi (1983)]{1983AA...123.121}Matteucci, F., \& Chiosi, C. 1983, \aap, 123, 121

\bibitem[Matteucci \& Tosi (1985)]{1985ARAA...217.391}Matteucci, F., \& Tosi, M. 1985, \mnras, 217, 391

\bibitem[Matteucci (1986)]{1986MNRAS...221.911}Matteucci, F. 1986, \mnras, 221, 911

\bibitem[Mirabel \& Wilson (1984)]{1984ApJ...277.92}Mirabel, I. F., $\&$ Wilson, A. S. 1984, \apj, 277,
92

\bibitem[Moran et al. (2001)]{2001ApJ...556.L75}Moran, E. C., et al. 2001, \apj, 556, L75

\bibitem[Mouhcine et al. (2005)]{2005MNRAS...362.1143}Mouhcine, M., Lewis, I., Jones, B., et al. 2005, \mnras, 362, 1143

\bibitem[Mouri \& Taniguchi (2002a)]{2002aApJ...565.786}Mouri, H., \& Taniguchi, Y. 2002a, \apj, 565, 786

\bibitem[Mouri \& Taniguchi (2002b)]{2002bApJ...566.L17}Mouri, H., \& Taniguchi, Y. 2002b, \apj, 566, L17

\bibitem[Mouri \& Taniguchi (2004)]{2004ApJ...605.144}Mouri, H., \& Taniguchi, Y. 2004, \apj, 605, 144

\bibitem[Nagao et al. (2003)]{2003AJ...26.1167}
Nagao, T., Murayama, T., Shioya, Y., \& Taniguchi, Y. 2003,
\aj, 126, 1167

\bibitem[Nagao et al. (2004)]{2004ApJ...128.109}Nagao, T., et al. 2004, \apj, 128, 109

\bibitem[Norman \& Scoville (1988)]{1988ApJ...332.124}Norman, C., \& Scoville, N. 1988, \apj, 332, 124

\bibitem[O'Halloran et al. (2006)]{2006ApJ...641.795}O'Halloran, B., Satyapal, S., \& Dudik, R, P. 2006, \apj, 641, 795

\bibitem[Oliva et al. (1999a)]{1999aAA...341.L75}Oliva, E., et al. 1999a, \aap, 341, L75

\bibitem[Oliva et al. (1999b)]{1999bAA...343.943}Oliva, E., et al. 1999b, \aap, 343, 943

\bibitem[Oliva et al. (1999c)]{1999cAA...350.9}Oliva, E., et al. 1999c, \aap, 350, 9;

\bibitem[Osterbrock (1993)]{1993ApJ...404.551}Osterbrock, D. E. 1993, \apj, 404, 551

\bibitem[Osterbrock \& Martel (1993)]{1993ApJ...414.552}
Osterbrock, D. E., \& Martel, A. 1993, \apj, 414, 552

\bibitem[Peeters et al. (2004)]{2004ApJ...613.986}
Peeters, E., Spoon, H. W. W., \& Tielens, A. G. G. M. 2004, \apj,
613, 986


\bibitem[Pilyugin (1993)]{1993AA...277.42} Pilyugin, L. S. 1993, \aap, 277, 42

\bibitem[Ramos Almeida et al. (2008)]{2008ApJ...680.L17}Ramos Almeida, C., et al. 2008, \apj, 680, L17

\bibitem[Renzini \& Voli (1981)]{1981AA...94.175}Renzini, A., \& Voli, M. 1981, \aap, 94, 175

\bibitem[Roussel et al. (2001)]{2001AA...372.427}Roussel, H., et al. 2001, \aap, 372, 427

\bibitem[Roussei et al. (2007)]{ApJ...669.959}Roussei, H., et al. 2007, \apj, 669, 959

\bibitem[Ruiz et al. (1994)]{1994ApJ...423.608}
Ruiz, M., Rieke, G. H., \& Schmidt, G. D. 1994, \apj, 423,
608

\bibitem[Rush et al. (1993)]{1993ApJS...89.1}
Rush, B., Malkan, M. A., \& Spinoglio, L. 1993, \apjs, 89, 1

\bibitem[Sanders \& Mirabel (1985)]{1985ApJ...298.L31}
Sanders, D. B., $\&$ Mirabel, I. F. 1985, \apj, 298, L31

\bibitem[Sanders et al. (1986)]{1986ApJ...305.L45}Sanders, D. B., et al. 1986, \apj, 305, L45

\bibitem[Sanders et al. (1988)]{1988ApJ...325.74}Sanders, D. B., et al. 1988, \apj, 325, 74

\bibitem[Sanders \& Mirabel (1996)]{1996ARAA34.749}Sanders, D. B.,\& Mirabel, I. F. 1996, \araa, 34, 749

\bibitem[Sanders et al. (2003)]{2003AJ...126.1607}Sanders, D. B., et al. 2003, \aj, 126, 1607

\bibitem[Schaerer \& Stasinska (1999)]{1999AA...345.L17}Schaerer, D., \& Stasinska, G. 1999, \aap, 345, L17

\bibitem[Schlegel et al. (1998)]{1998APJ...500.525}Schlegel, D. J., Finkbeiner, D. P., \& Davis, M. 1998, \apj, 500,525

\bibitem[Shi et al. (2007)]{2007ApJ...669.841}Shi, Y., et al. 2007, \apj, 669, 841

\bibitem[Shi et al. (2010)]{2010ApJ...714.115}Shi, Y., et al. 2010, \apj, 714, 115

\bibitem[Shu et al. (2007)]{2007ApJ...657.167}Shu, X. -W., et al. 2007, \apj, 657, 167

\bibitem[Smith et al. (2004)]{2004ApJS...154.199}Smith, J. D. T., et al. 2004, \apjs, 154, 199

\bibitem[Smith et al. (2007)]{2007ApJ...656.770}Smith, J. D. T., et al. 2007, \apj, 656, 770

\bibitem[Sosa-Brito et al. (2001)]{2001ApJS...136.61}
Sosa-Brito, R. M., et al. 2001, \apjs, 136, 61

\bibitem[Spinoglio \& Malkan (1989)]{1989ApJ...342.83}
Spinoglio, L., \& Malkan, M. A. 1989, \apj, 342, 83

\bibitem[Spinoglio \& Malkan (1992)]{1992ApJ...399.504}
Spinoglio, L., \& Malkan, M. A. 1992, \apj, 399, 504

\bibitem[Spoon et al. (2007)]{2007ApJ...654.L49}Spoon, H. W. W., et al. 2007, \apj, 654, L49

\bibitem[Stasi$\rm \acute{n}$ska et al. (2006)]{2006MNRAS371.972}
Stasi$\rm \acute{n}$ska G., et al. 2006. \mnras, 371, 972

\bibitem[Sternberg \& Dalgarno (1995)]{1995ApJS...99.565}
Sternberg, A., \& Dalgarno, A. 1995, \apjs, 99, 565

\bibitem[Storchi-Bergmann et al. (2001)]{2001ApJ...559.147}Storchi-Bergmann, T., et al. 2001, \apj, 559, 147

\bibitem[Sturm et al. (2002)]{2002AA...393.821}Sturm, E., et al. 2002, \aap, 393, 821

\bibitem[Sturm et al. (2005)]{2005ApJ...629.L21}Sturm, E., Schweitzer, M., Lutz, A.,
Contursi, A., Genzel, R., Lehnert, M. D., Tacconi, L. J., Veilleux,
S., Rupke, D. S., \& Kim, D. C. 2005, \apj, 629, L21

\bibitem[Sturm et al. (2006)]{2006ApJ...653.L13}Sturm, E. et al., 2006, \apj, 653, L13

\bibitem[Surace et al. (2004)]{2004AJ...127.3235}Surace, J. A., et al. 2004, \aj, 127, 3235

\bibitem[Taniguchi et al. (1999)]{1999ApJ...514.L9}
Taniguchi, Y., Ikeuchi, S., \& Shioya, Y. 1999, \apj, 514, L9

\bibitem[Thornley et al. (2000)]{2000ApJ...539.641}Thornley, M. D., et al. 2000, \apj, 539, 641

\bibitem[Tommasin et al. (2008)]{2008ApJ...676.836}Tommasin, S., et al. 2008, \apj, 676, 836

\bibitem[Tommasin et al. (2010)]{2010ApJ...709.1257}Tommasin, S., et al. 2010, \apj, 709, 1257

\bibitem[Toomre \& Toomre (1972)]{1972ApJ...178.623}
Toomre, A., \& Toomre, J. 1972, \apj, 178, 623

\bibitem[Toomre (1977)]{1977ARAA...15.437}Toomre, A. 1977, \araa, 15, 437

\bibitem[Tran (2001)]{2001ApJ...554.L19}Tran, H. D. 2001, \apj, 554, L19

\bibitem[Tran (2003)]{2003ApJ...583.632}Tran, H. D. 2003, \apj, 583, 632

\bibitem[Tran (2010)]{2010ApJ...711.1174}Tran, H. D. 2010, \apj, 711, 1174

\bibitem[Tran et al. (2000)]{2000AJ...120.562}
Tran, H. D., Cohen, M. H., \& Villar-Martin,M. 2000, \aj, 120, 562

\bibitem[Vaceli et al. (1997)]{1997AJ...114.1345}
Vaceli, M. S., Viegas, S. M., Gruenwald, R., \& De
Souza, R. E. 1997, \aj, 114, 1345

\bibitem[van Dokkum (2005)]{2005AJ...130.2647}van Dokkum, P. G. 2005, \aj, 130, 2647

\bibitem[Veilleux \& Osterbrock (1987)]{1987ApJS...63.295}
Veilleux, S., \& Osterbrock, D. E. 1987, \apjs, 63, 295

\bibitem[Veilleux et al. (2009)]{2009ApJS...182.628}Veilleux, S., et al. 2009, \apjs, 182, 628

\bibitem[Verma et al. (2003)]{2003AA...403.829}Verma, A., Lutz, D., Sturm, E.,
Sternberg, A., Genzel, R., \& Vacca. 2003, \aap, 403, 829

\bibitem[Vigroux et al. (1987)]{1987AA...172.15}Vigroux, L., Stasinska, G., \& Comte, G. 1987, \aap, 172, 15

\bibitem[Voit (1992)]{1992ApJ...399.495}Voit, G. M. 1992, \apj, 399, 495


\bibitem[Wang  \& Zhang (2007)]{2007ApJ...660.1072}
Wang, J.-M., \& Zhang, E.-P. 2007, \apj, 660, 1072

\bibitem[Weedman et al. (20050]{2005ApJ...633.706}Weedman, D. W., et al. 2005, \apj, 633, 706

\bibitem[Wu et al. (2006)]{2006ApJ...639.157}Wu, Y. -L., et al. 2006, \apj, 639, 157

\bibitem[Wu et al. (2009)]{2009ApJ...701.658}Wu, Y. -L., et al. 2009, \apj, 701, 658

\bibitem[Wu et al. (2011)]{2011ApJ...730.121}
Wu, Y. -Z., et al. 2011, \apj, 730, 121

\bibitem[Young et al. (1986)]{1986ApJ...304.443}
Young, J. S., Schloerb, F. P., Kenney, J., \& Lord , S. D. 1986,
\apj, 304, 443

\bibitem[Young et al. (19960]{1996MNRAS281.1206}Young, S., et al. 1996, \mnras, 281, 1206

\bibitem[Yuan et al. (2010)]{2010ApJ...709.884}Yuan, T. -T., et al. 2010, \apj, 709, 884

\bibitem[Zhang \& Wang (2006)]{2006ApJ...653.137}
Zhang, E. -P., \& Wang, J. -M. 2006, \apj, 653, 137

\bibitem[Zhang et al. (2008)]{2008ApJ...685.L109}
Zhang, K., Wang, T. -G., Dong, X. -B., \& Lu, H. -L.
2008, \apj, 685, L109


\end{thebibliography}
\end{document}